\documentclass[letterpaper]{ar-1col}

\usepackage[margin=1.5in]{geometry}

\def\la{\mathrel{\hbox{\rlap{\hbox{\lower4pt\hbox{$\sim$}}}{\raise2pt\hbox{$<$}}}}}
\def\ga{\mathrel{\hbox{\rlap{\hbox{\lower4pt\hbox{$\sim$}}}{\raise2pt\hbox{$>$}}}}}

\usepackage{natbib}
\newcommand\aapr{A\&AR}
\newcommand\aj{AJ}
\newcommand\araa{ARA\&A}
\newcommand\apj{ApJ}
\newcommand\apjl{ApJ}
\newcommand\apjs{ApJS}
\newcommand\aap{A\&A}
\newcommand\pasj{PASJ}

\newcommand\mnras{MNRAS}
\newcommand\nat{Nature}

\jname{Xxxx. Xxx. Xxx. Xxx.}
\jvol{AA}
\jyear{YYYY}
\doi{10.1146/((please add article doi))}

\begin{document}

\markboth{Kaaret, Feng, \& Roberts}{Ultraluminous X-Ray Sources}

\title{Ultraluminous X-Ray Sources}

\author{Philip Kaaret,$^1$ Hua Feng,$^2$ and \newline Timothy P.~Roberts$^3$
\affil{$^1$Department of Physics and Astronomy, University of Iowa, Iowa City, 52242, USA; email: philip-kaaret@uiowa.edu}
\affil{$^2$Department of Engineering Physics and Center for Astrophysics, Tsinghua University, Beijing, 100084, China}
\affil{$^3$Centre for Extragalactic Astronomy, Department of Physics, Durham University, South Road, Durham DH1 3LE, UK}}

\begin{abstract}

We review observations of ultraluminous X-ray sources (ULXs). X-ray spectroscopic and timing studies of ULXs suggest a new accretion state distinct from those seen in Galactic stellar-mass black hole binaries. The detection of coherent pulsations indicates the presence of neutron-star accretors in three ULXs and therefore apparently super-Eddington luminosities. Optical and X-ray line profiles of ULXs and the properties of associated radio and optical nebulae suggest that ULXs produce powerful outflows, also indicative of super-Eddington accretion. We discuss models of super-Eddington accretion and their relation to the observed behaviors of ULXs. We review the evidence for intermediate mass black holes in ULXs. We consider the implications of ULXs for super-Eddington accretion in active galactic nuclei, heating of the early universe, and the origin of the black hole binary recently detected via gravitational waves.

\end{abstract}

\begin{keywords}
accretion physics, black holes, neutron stars, X-ray binaries, X-rays: galaxies
\end{keywords}

\maketitle

\tableofcontents

\section{INTRODUCTION}

The most basic observational property of an X-ray source is its flux which, under the assumption of a radiation pattern, can be translated to a luminosity. Ultraluminous X-ray sources (ULXs) are non-nuclear objects found in external galaxies that appear to be very luminous under the assumption of isotropic emission. The majority of ULXs are thought to be X-ray binaries powered by accretion onto a compact object and their luminosities are comparable to or above the Eddington luminosity of stellar black holes. Thus, ULXs offer a means to study either accretion near and above the Eddington limit, if the compact objects are stellar mass black holes, or unusually massive black holes \citep{Colbert1999}.

The key physical properties of an ultraluminous X-ray binary are the mass, spin, and nature (black hole or neutron star) of the compact object, the nature of the companion star, and the orbital separation and eccentricity. These are related via the geometry and dynamics of the accretion flow, particularly the accretion rate, to the observed X-ray, optical/UV/IR, and radio emission from the system. In the following, we first review the observational properties of ULXs including X-ray spectroscopy and timing and studies of their multiwavelength counterparts. We then discuss models for ULXs with emphasis on super-Eddington accretion and review the evidence for unusually massive black holes in ULXs. We conclude with a discussion of the implications of the knowledge gleaned from studies of ULXs for other topics in astrophysics.

\begin{figure}[tb]
\includegraphics[width=3in]{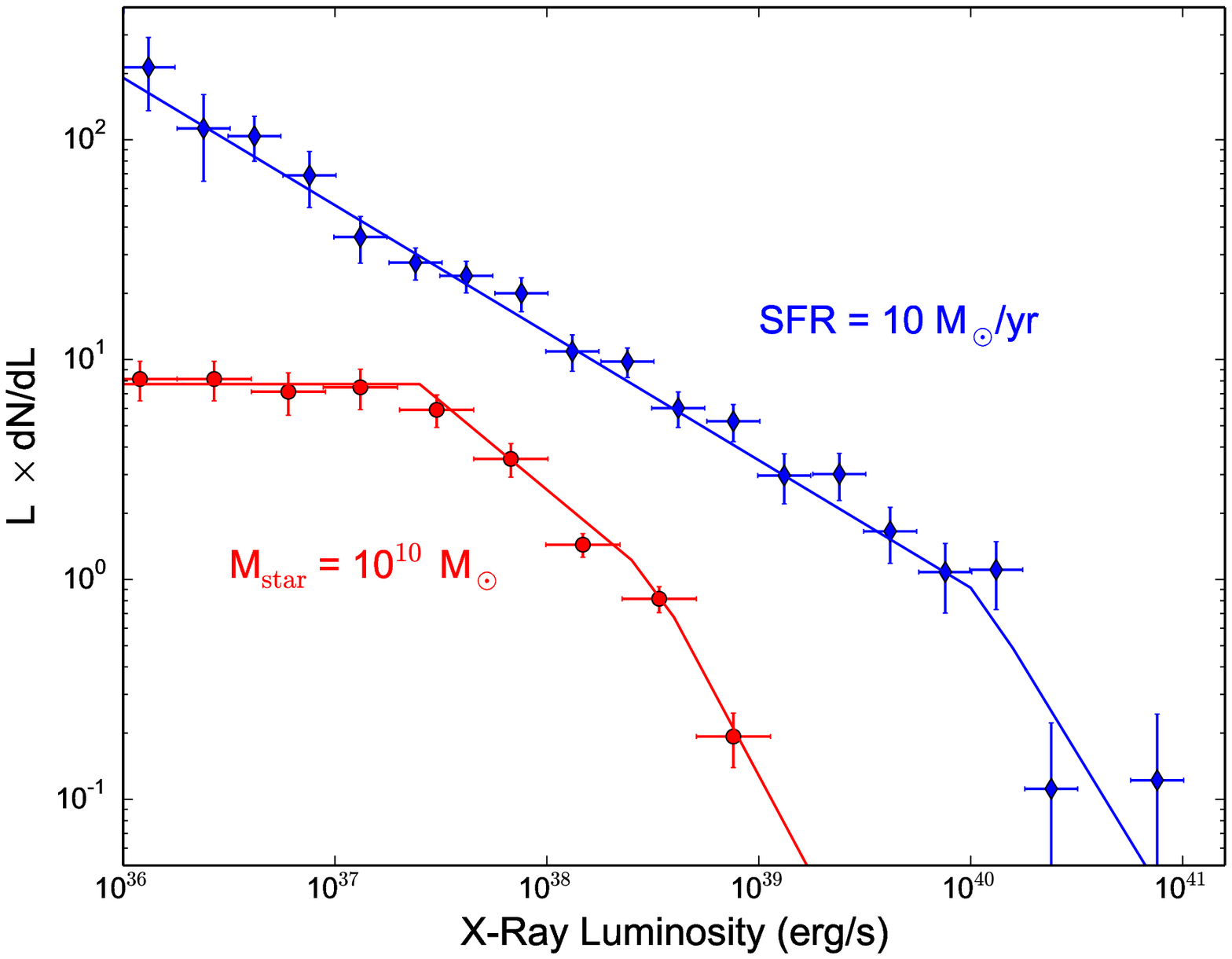}
\caption{Point-source X-ray luminosity functions for star-forming (blue) and elliptical (red) galaxies following \citet{Gilfanov2004} with data from \citet{Mineo2012}.} 
\label{fig:xlf}
\end{figure}

\subsection{X-ray luminosities}
\label{sec:xlf}

The {\it Einstein} observatory obtained the first resolved images of external galaxies that enabled the discovery of the first ULXs \citep{Long1981,Fabbiano1989}. ULXs are defined as point-like sources within or near the optical extent of a host galaxy, but away from the nucleus in order to exclude accreting supermassive black holes. If defined strictly on the observed flux, the ULX population contains a small number of bright supernovae and supernova remnants. Young rotation-powered pulsars may theoretically reach sufficiently high luminosities  \citep{Perna2004,Medvedev2013}, but no example has been identified. X-ray binaries can be distinguished by their irregular variability and relatively dim optical counterparts. This review focuses on ultraluminous X-ray binaries \citep{Feng2011}.

Several different threshold luminosities have been used in classifying ULXs. When introducing the term `ultraluminous compact X-ray sources', \citet{Makishima2000} used the Eddington limit for a $1.4 M_{\odot}$ neutron star. A common definition in use today is $1 \times 10^{39} \, \rm erg \, s^{-1}$ which is convenient in cgs units and tends to be used for population studies as it produces larger samples. Another commonly used definition is $3 \times 10^{39} \rm \, erg \, s^{-1}$ which closely corresponds to the Eddington luminosity for a 20~$M_{\odot}$ black hole, more massive than any stellar black hole observed within the Milky Way \citep{Remillard2006}, and use of this definition is an attempt to establish ULXs as a class physically distinct from stellar-mass X-ray binaries found in the Milky Way.

The X-ray luminosity functions (XLFs) of actively star-forming and old elliptical galaxies are shown in Figure~\ref{fig:xlf}. The XLF for star-forming galaxies has a break at (1--2$)\times 10^{40} \rm \, erg \, s^{-1}$ \citep{Mineo2012,Swartz2011}, while that for ellipticals has a break at $(5 \pm 2) \times 10^{38} \rm \, erg \, s^{-1}$ \citep{Kim2010,Gilfanov2004}. The XLFs are plotted assuming isotropic luminosity. Beaming, which reaches a factor of 2 for a thin accretion disk and potentially higher for other accretion geometries (Section~\ref{sec:timing}), could shift an observed luminosity break above the physical luminosity at which the break occurs. Hence, the break in the elliptical XLF may correspond to the Eddington luminosity for neutron stars and may suggest that there are few black hole X-ray binaries in such galaxies. The original definition of a ULX in terms of the Eddington limit for a $1.4 M_{\odot}$ neutron star, after allowing for beaming from a thin disk, is justified by the break in the XLF of elliptical galaxies.

The break in the star-forming XLF would correspond to objects 20 times as massive. However, the lack of a break near the neutron star Eddington luminosity in the star-forming XLF, together with the fact that neutron star binaries are estimated to be 10--50 times more numerous than black hole binaries \citep{Belczynski2009}, calls into question if either XLF break is related to an Eddington luminosity. ULXs with luminosities up to the break therefore appear to be part of the same population as standard X-ray binaries.

Sources with luminosities above the break, $L_X > 2 \times 10^{40} \rm \, erg \, s^{-1}$, could represent a new class of objects. Interestingly, extreme objects with luminosities above the break do exist. The first example was found in M82 \citep{Kaaret2001,Matsumoto2001} and the term `hyperluminous X-ray source' (HLX) was first used in \citet{Matsumoto2003}. HLXs are defined as sources reaching a peak luminosity of at least $10^{41} \rm \, erg \, s^{-1}$ \citep{Gao2003} and ones with luminosities up to $10^{42} \rm \, erg \, s^{-1}$ have been found and are discussed below. These sources may be the best candidates for intermediate mass black holes (IMBHs).

The first systematic surveys of nearby galaxies for X-ray binaries were done using the {\it R\"ontgensatellit} High Resolution Imager \citep{Colbert1999,Roberts2000}. The {\it Chandra} X-ray observatory greatly increased the sensitivity and, critically, enabled accurate source localization. Background active galactic nuclei (AGN) and foreground stars often masquerade as ULXs \citep[e.g.][]{Gutierrez2013}. These contaminates can be revealed by bright, point-like optical counterparts if accurate positions are available. 

\citet{Swartz2011} identified 107 ULX candidates ($L_X > 1 \times 10^{39} \, \rm erg \, s^{-1}$) in a complete sample of galaxies within 14.5~Mpc and with masses above $10^{7.5} M_{\odot}$, giving a local ULX density of one per 57 Mpc$^3$. Their XLF predicts at most one HLX ($L_X > 1 \times 10^{41} \, \rm erg \, s^{-1}$) within 100~Mpc, requiring a distinct population to explain the observed HLXs. The largest samples of ULXs and HLXs have come from {\it XMM-Newton} due to its larger field of view and collecting area. \citet{Walton2011} present a catalog of 470 ULX candidates in 238 nearby galaxies with an estimated contamination of $\sim$20\%. The X-ray binary population of a galaxy is linearly correlated with both its star formation rate and stellar mass \citep{Colbert2004}. Dominance of the former in star forming galaxies suggests that their ULX population is young with ages of tens of Myr, while dominance of the latter in elliptical galaxies suggest their ULX population is old with ages of Gyr.

ULXs are preferentially associated with star-forming galaxies, as can be inferred from the XLFs, and also with the star forming regions within galaxies \citep{Swartz2009}. Production of ULXs, per unit star formation rate, increases at sub-solar metallicities \citep{Mapelli2010} and is enhanced by a factor of $7 \pm 3$ in number \citep{Prestwich2013} and $11.5 \pm 2.7$ in total X-ray luminosity \citep{Brorby2014} at very low metalliticies, $Z/Z_{\odot} < 0.1$. The maximum stellar black hole mass also increases at low metallicity, which could decrease the need for super-Eddington accretion in ULXs \citep{Zampieri2009}.

\subsection{Comparison with Galactic X-ray binaries}
\label{sec:gbhb}

X-ray binaries within the Milky Way have been extensively studied and are an important benchmark in studying ULXs \citep{Remillard2006}. Galactic black hole X-ray binaries (GBHBs) exhibit pronounced variability with the X-ray flux changing by a factor as large as $10^7$ in some cases \citep{Remillard2006}.  The majority of GBHBs have low-mass companion stars and are transient objects that are usually in a quiescent state and flare to high luminosities, sometimes approaching the Eddington limit, for weeks to months. In contrast, most ULXs are persistent for years or decades. ULXs in elliptical galaxies typically have low levels of variability \citep{Feng2006}. ULXs in star-forming galaxies are variable by factors often reaching 10 \citep{Kaaret2009}. Variability by large factors, exceeding 100, is unusual, but seen in some ULXs, \citep[e.g.][]{Bachetti2014}.

GBHBs exhibit a variety of X-ray spectral/timing states thought to be connected with the accretion rate and geometry \citep[e.g.,][]{Remillard2006}. The X-ray spectra are commonly modeled as the sum of a multicolor disk blackbody component thought to arise from the accretion disk and a power-law or Comptonization component thought to arise from a corona. Power spectra of the X-ray variability typically show broadband noise extending from low frequencies up to about 10~Hz and sometimes exhibit quasiperiodic oscillations (QPOs). A typical transient GBHB exhibits one or more spectral/timing states during outburst. Outbursts typically begin and end in the {\it hard} state which shows a power-law X-ray spectrum with a hard photon index ($1.4 < \Gamma < 2.1$) and strong variability on short timescales \citep{Belloni2010}. The hard state is usually accompanied by a compact radio jet. During the outburst, the source may enter into the {\it thermal} state, the {\it steep power law} state, or one or more intermediate states that are less clearly defined. In the {\it thermal} state, the accretion disk dominates the spectrum, fast variability is weak, and the compact radio jet is absent. The {\it steep power law} state or {\it very high} state \citep{Miyamoto1991} has strong emission from both the disk and a power-law component with a steep photon index, $\Gamma \sim 2.5$ and is seen at luminosities approaching the Eddington limit. This state often exhibits QPOs, including ones at high frequency.

Spectral variability provides a strong test of the canonical GBHB spectral model. In particular, if the disk component truly represents a standard thin accretion disk extending all the way to the innermost stable circular orbit (ISCO) around the black hole, then the disk should have a constant inner radius. In that case, the bolometric disk luminosity, $L_{\rm disk}$, and the disk inner temperature, $T_{\rm in}$, should be related as $L_{\rm disk} \propto T^4$. This is demonstrated with good accuracy in the thermal state of GBHBs \citep[e.g.][]{Gierlinski2004}. The radius of the ISCO is a function of only the black hole mass and spin, so the normalization of the relation can be used to infer the black hole mass given a spin value.

It is interesting to ask how, or even if, ULXs relate to these sub-Eddington spectral states. However, before we do, we need to introduce an important caveat.  While it is common to directly compare the shape of ULX spectra and those of GBHBs, most observational results for the two classes are subject to a bandpass mismatch.  Observational data for GBHBs has predominantly been gathered by instruments designed for Galactic science, primarily gas proportional counters, such as the Proportional Counter Array on the {\it Rossi X-ray Timing Explorer} (RXTE), that have limited sensitivity below 3 keV but large collecting area at high energies up to $\sim$25~keV. In contrast, ULXs are generally observed by observatories that are more specifically designed for extra-galactic science, with CCD detectors and focusing optics, and so operate in the $\sim$0.5--10~keV regime. Observations of ULXs therefore go much softer than those of BHBs, and so are sensitive to soft features that are not observed in many BHB datasets. Similarly most ULX datasets, before the advent of NuSTAR, were not sensitive to the hard spectral components seen in GBHB data.  This should be borne in mind when interpreting ULX data.

\section{X-RAY SPECTRA}
\label{sec:xrayspectra}


\subsection{Simple Models}

Our first views of ULX spectra in the 0.5--10~keV band, where there is sufficient bandpass to start discriminating between physical models, emerged from the {\it ASCA\/} mission in the late 1990s \citep{Colbert1999}. These data showed that many ULXs have spectra that appear to be dominated by a single spectral component that is somewhat convex in shape, and so can be fitted with accretion disk models \citep{Makishima2000}. These disks are both hotter (with $kT_{\rm in} \approx 1.1 - 1.8$ keV) and more luminous than those seen in GBHBs, and \citet{Makishima2000} suggested this is a consequence of ULXs hosting larger stellar mass black holes, up to $\sim 100 M_{\odot}$ in order to remain sub-Eddington, that are rapidly rotating, causing the accretion disk to move closer to the black hole and heat up.  Some ULXs also show {\it ASCA\/} spectra that appear power-law dominated in some epochs \citep[e.g. two ULXs in IC 342,][]{Kubota2001}. While these can be interpreted as a high luminosity manifestation of the hard state, \citet*{Kubota2002} note that a strongly Comptonized disk spectrum provides a better physical explanation, consistent with objects in the steep power-law/very high state.

Early ULX studies with both {\it Chandra\/} and {\it XMM-Newton\/} also largely used similar single component models to fit ULX spectra \citep[e.g.][]{Foschini2002,Roberts2002}. However, it quickly became apparent from deeper observations, particularly with {\it XMM-Newton\/} but also with {\it Chandra\/}, that single component models in the form of power-laws or simple accretion disk models such as the standard multi-color disk blackbody (DBB) are statistically excluded in virtually all cases with sufficiently high spectral quality. The spectra required either the inclusion of a second component or more complex single component models to obtain adequate fits. We focus our discussion below on these models. We note that where the data quality is low, e.g.\ in short exposures or distant galaxies, single component models cannot be statistically excluded; however, this should not be taken as evidence that the ULX spectra have that physical origin, e.g.\ an acceptable hard power-law fit alone is not sufficient evidence to indicate the X-ray hard state.

\subsection{Continuum Models}
\label{sec:contspec}

The first good quality {\it XMM-Newton\/} and {\it Chandra\/} spectra were fitted with the same two-component model used in GBHBs: a soft accretion disk and a power-law. When fitted to many ULX spectra this results in very cool inner disk temperatures, $kT_{\rm in}$ around 0.1--0.3~keV, and relatively hard power-laws, $\Gamma \sim 1.5$--3 \citep[e.g.,][]{Kaaret2003,Miller2003,Cropper2004}. Crucially, if this disk temperature represents that of the innermost stable circular orbit around the black hole, then we expect that $T_{\rm in} \propto M^{-0.25}$ where $M$ is the black hole mass, for fixed Eddington ratio and black hole spin. Scaling from the typical temperatures of GBHBs in the thermal dominant state of $\sim$1~keV, this would imply that IMBHs with masses $\sim 10^3 M_{\odot}$ are present in many ULXs. However, in the ULX spectra, the power-law dominates above 1~keV, which is in sharp contrast to GBHB spectra in the thermal dominant state where the disk dominates in the 2-10~keV band. There are other technical issues in that continuation of the power-law to lower energies is unphysical, the absorption column density is usually strongly correlated with the disk temperature and flux complicating estimation of the latter, and the fits often have a second local minimum with a hotter disk ($kT_{\rm in} =$~0.8--2.5~keV) and a steep power-law ($\Gamma =$~2--4.5).

\begin{figure}[bt]
\includegraphics[width=4.5in]{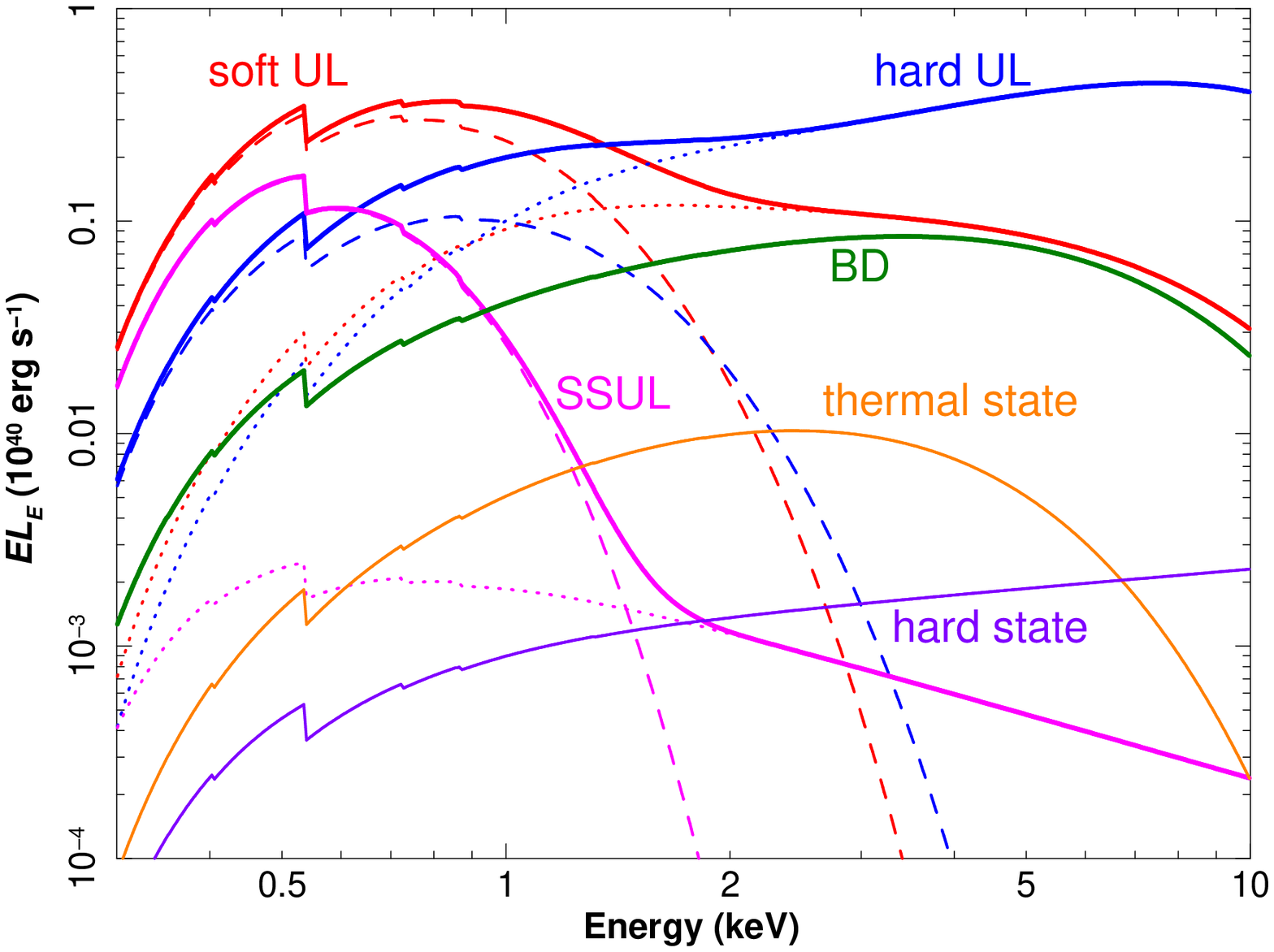}
\caption{A comparison of different ULX and GBHB spectra. \newline
{\bf Ultraluminous (UL) --} spectra are distinctly two-component, with a soft excess and a hard component that shows a turn-over at high energies. UL spectra are classified as hard (blue) versus soft (red) according to which component peaks higher in energy density space. When modeled as a DBB (dashed lines) plus corona (dotted lines), the disk temperature is typically $0.15-0.3$~keV while the corona temperature is $\sim 1.5-3$~keV. The corona is optically thick with $\tau > 6$. Both hard and soft UL spectra are shown with a total intrinsic $L_{\rm X} = 10^{40} \rm \,erg \, s^{-1}$. \newline 
{\bf Broadened disk (BD) --} spectra (green) appear accretion-disk like, but are too broad to be fitted by simple accretion disk models such as the standard multicolor disk blackbody; instead they tend to be well fitted by the $p$-free disk model with $kT_{\rm in} \sim 1-2.5$~keV and $p \sim 0.6$. If modeled as a DBB plus power-law, the power-law component acts to broaden the disk spectrum by emerging from beneath the disk spectrum at one or both ends of the spectrum. The spectrum shown is a $p$-free disk model with $L_{\rm X} = 2 \times 10^{39} \rm \, erg \, s^{-1}$.  \newline
{\bf Supersoft ultraluminous (SSUL) --} spectra (magenta) are dominated by a cool blackbody component, with $kT \sim 0.1$~keV, producing over 90\% of the intrinsic flux in the 0.3--10~keV band. A hard component, with $\Gamma = 2-4$ if modeled as a power-law, is weak but evident in high quality data. The spectrum shown is a blackbody plus power-law model with $L_{\rm X} = 3 \times 10^{39} \rm \, erg \, s^{-1}$. \newline
The other curves show standard GBHB spectra: thermal dominant (orange) represented by a DBB with $kT_{\rm in} = 1$~keV and $L_X = 2 \times 10^{38} \rm \, erg \, s^{-1}$; and hard (purple) shown as a power-law with $\Gamma = 1.7$ and $L_X = 5 \times 10^{37} \rm \, erg \, s^{-1}$. The absorption is $1 \times 10^{21} \rm \, atom \, cm^{-2}$ in all cases and is responsible for the edge features below 1~keV.}
\label{fig:ULXspecs}
\end{figure}

Perhaps more importantly, higher quality {\it XMM-Newton\/} data revealed that the hard emission is not a power-law; it shows curvature in the 2--10~keV band \citep[e.g.][]{Roberts2005,Feng2005}. In the highest quality {\it XMM-Newton\/} data this curvature appears statistically significant in the majority of cases \citep{Stobbart2006,Gladstone2009}. This curvature has been spectacularly highlighted by recent {\it NuSTAR\/} observations, with its extended hard bandpass clearly showing the curvature extending above 10 keV \citep[e.g.,][]{Bachetti2013,Walton2014}. 

The curvature is well described by a break at energies of 2--7~keV \citep{Gladstone2009}, well below the breaks seen in GBHB which are typically $\sim$60~keV or above \citep{McClintock2006}. Hence the canonical GBHB X-ray spectral model does not apply to ULXs and interpretation of the cool component as the disk of an IMBH accreting in a sub-Eddington state appears very questionable. Indeed, the general appearance of the ULX spectra, a hard component displaying curvature and a soft excess, does not directly map on to any of the standard sub-Eddington states. The two-component spectra are almost always seen at luminosities above $3 \times 10^{39} \rm \, erg \, s^{-1}$. It was therefore proposed that this may represent a super-Eddington {\it ultraluminous} (UL) state \citep{Gladstone2009}.

A physically-motivated two-component model, based on the physical interpretation of the canonical GBHB model, is a DBB plus Comptonized corona. In these the disk temperature is again cool, $kT_{\rm in}$ around 0.1--0.3~keV, but in contrast to GBHBs where the coronae are hot and thin (with electron temperature $kT_{\rm e} \sim 100$ keV and optical depth $\tau \la 1$), the coronae of ULXs appear cool and thick with  $kT_{\rm e} \sim$1--2~keV and $\tau > 6$. The relative flux of the disk and corona component varies. Those in which the hot component dominates are `hard UL' spectra, while the cool component dominates in `soft UL' spectra \citep{Sutton2013_435}. Illustrative examples are shown in Fig.~\ref{fig:ULXspecs}.

These fits led to the suggestion that the inner accretion disks of ULXs are shrouded by an optically-thick corona leaving only the outer regions of the disk visible as the soft excess. A model based on this scenario, including energy coupling of the corona and disk, was used by \citet{Gladstone2009} to recover temperatures of the underlying accretion disk of $\sim 1$~keV in most (but not all) cases, consistent with stellar-mass black holes. However, two component spectra such as those observed might also be expected from supercritical accretion disks (Section~\ref{sec:bhsupereddmodels}).

Some high quality ULX spectra appear more like a single, broad continuum than two distinct components, and can be well fitted with the $p$-free disk model, where the radial temperature profile of the disk is allowed to vary as $T(R) \propto R^{-p}$. These show $p \approx0.5-0.6$, as expected if the disk becomes radiatively inefficient due to the advection of radiation in its interior. This exponent is lower than the $p = 0.75$ of a standard thin disk which broadens the range of temperatures contributing to the emission and therefore the resulting energy spectrum. These spectra are referred to as `broadened disk' (BD). BD spectra are seen to dominate below luminosities of $3 \times 10^{39} \rm \, erg \, s^{-1}$. They are also seen from several objects with luminosities well above $10^{40} \rm \, erg \, s^{-1}$, for example ULXs in NGC 5907 \citep{Sutton2013_434} and NGC 5643 \citep{Pintore2016}. It is unclear whether this resemblance to lower-luminosity BD spectra is due to the same underlying physical processes or represents a different spectral regime.  


Beyond confirming the spectral curvature at high energies found with {\it XMM-Newton\/}, several simultaneous {\it NuSTAR\/}/{\it XMM-Newton\/} spectra hint at the presence of an additional hard component. The strongest detection is in Ho II X-1 \citep{Walton2015} and the emission can be modeled as an optically thin, hot or non-thermal electron scattering corona with $\Gamma \sim 3$, as seen in GBHBs in the steep power-law/very high state. If this is a corona then it signifies that the hard component of the $0.3-10$ keV spectrum cannot be a corona; instead it is most likely emission from the inner regions of the accretion disk, see Section~\ref{sec:bhsupereddmodels}.


\subsection{Supersoft ULXs}
\label{sec:ssulx}

There is a sub-class of ULXs for which most of the flux is emitted below 1 or 2 keV. The spectra of these very soft or supersoft ULXs can be decomposed into a dominant blackbody component, with a temperature typically of 100~eV, and a minor power-law component that extends to high energies with $\Gamma \sim$2-4. These supersoft ULXs (SSULXs) are also referred to as ultraluminous supersoft X-ray sources or ULSs \citep{Liu2008}. Only a handful of these sources have been well studied. The observed luminosities of these sources are mostly around or below $10^{39} \rm ~erg~s^{-1}$, although their inferred luminosities once absorption is removed are often much higher \citep{Urquhart2016}.

\subsection{Absorption}

We have so far dealt with ULX spectra as continua. However, all are subject to absorption from cold, neutral material along our line of sight to the ULX, whether it lies in our Galaxy, the host galaxy of the ULX, or is intrinsic to the ULX itself. However, establishing the correct absorption columns to ULXs using continuum modeling is fraught with uncertainty, as the measured absorption is strongly dependent upon the continuum model used, e.g.\ power-laws return far higher columns than disk models due to the absorption being the sole cause of the low-energy downturn in spectra modeled by a power-law.  

In general, most models show that ULXs are not subject to very high columns, with typical absorption columns of $\sim$(1--3)$\times 10^{21} \rm \, cm^{-2}$ in excess of Galactic foreground absorption, although there are a minority of objects with higher absorption \citep[e.g.\ IC 342 X-2 with a column above $10^{22} \rm ~cm^{-2}$;][]{Sutton2013_435}. \citet{Winter2007} suggest that these columns are not local to the ULX, but instead consistent with the line-of-sight in the host galaxy, based on early {\it XMM-Newton\/} data. However, some ULXs do show evidence that their column varies which puts this is doubt \citep{Middleton2015_454}. There are also additional uncertainties caused by the metallicity of the neutral absorber \citep[e.g.][]{Kaaret2004}.

\subsection{Spectral Variability}
\label{sec:spec_var}

Early {\it ASCA\/} spectroscopy of ULXs revealed that individual ULXs can display different spectra in different observational epochs \citep[e.g.,][]{Kubota2001}. Perhaps the simplest way of tracking this spectral variability with time is through X-ray color analyses. \citet{Roberts2006} monitored NGC 5204 X-1 with {\it Chandra} and found that its colors demonstrated spectral hardening with increased luminosity. Subsequent studies have concentrated mainly on the large monitoring datasets for individual ULXs acquired by {\it Swift}, but the low statistics of individual {\it Swift} datasets do not place strong constraints on X-ray colors, and so studies tend not to show significant trends in color changes with flux \citep{Kaaret2009}, although there can be some scatter between individual datasets at similar count rates \citep{Luangtip2016}.  


However, X-ray spectral data from large telescopes are usually good enough that it is more useful to directly contrast the spectra obtained in different epochs.  Both \citet{Kajava2009} and \citet{Feng2009} did this for a small number of ULXs observed on multiple occasions by {\it XMM-Newton}. Both found correlated $\Gamma - L_{\rm X}$ behavior for some ULXs that could be modeled as power-laws, though the direction of this behavior changed in different ULXs: some softened, and others hardened, at higher luminosities. \citet{Kajava2009} found that four ULXs that could be fitted with single component disk-like spectra showed positive disk luminosity versus temperature evolution, as seen from GBHBs. 

In contrast, \citet{Feng2007} studied the evolution of disk luminosity and temperature in the (cool) DBB plus power-law model for NGC 1313 X-2 and found that the disk luminosity decreases with temperature, in strong contrast to $L_{\rm disk} \propto T_{\rm in}^{4}$ relation expected for an accretion disk with a fixed inner radius. \citet{Kajava2009} studied a larger ULX sample and found that $L_{\rm disk} \propto T_{\rm in}^{-3.5}$. However, this result is disputed by \citet{Miller2013}, who find that the soft components of some ULXs do vary close to the predicted $L \propto T^4$ for standard disks. They assume a constant absorption column, which could strongly influence the fitted relationship, while some ULXs do show evidence for varying column as noted above. Interestingly, \citet*{Luangtip2016} find that for Ho IX X-1 at least, the positive $L$--$T$ relationship is an artifact of fitting a 2-component model as the spectrum transitions from a two-component to a disk-like spectrum at its highest luminosities. Indeed, the cooling of the soft component with increased luminosity is strongly confirmed in SSULXs \citep{Urquhart2016} which offer the advantage of a relatively clean view of the soft component given the weakness of the hard component. The blackbody radius is found to be inversely scaled with the blackbody temperatures, in a relation close to $R_{\rm bb} \propto T_{\rm bb}^{-2}$, for both individual sources \citep{Soria2016,Feng2016} and the whole population \citep{Urquhart2016}. 

There is some evidence that ULXs can change between spectral forms \citep{Sutton2013_435}. Some objects with hard UL spectra at lower luminosities appear to transition to BD spectra at their peak luminosities, for example NGC 1313 X-2 \citep{Pintore2012,Middleton2015_447} and Ho IX X-1 \citep{Walton2014,Luangtip2016}. NGC 247 X-1 may link SSULXs and broad-band ULXs, as it was found to make a transition from a supersoft state to a state with substantial hard X-ray emission (comparable flux from the power-law and blackbody components in 0.3-8~keV band), similar to the spectrum of some normal ULXs \citep{Feng2016}. There is also evidence for spectral degeneracy, with different spectra seen at the same luminosity in some ULXs \citep[e.g.][]{Grise2010,Vierdayanti2010,Marlowe2014}.

\subsection{X-ray evidence for outflows}
\label{sec:xrayoutflows}

Low-energy line-like spectral residuals seen in ULXs have been attributed to incorrectly modeled low metallicity absorption \citep{Cropper2004,Goad2006}. However, this may not be the only cause for low energy spectral residuals. Early ULX analyses often attributed these residuals to possible confusion with supernova remnants and/or a hot component of the interstellar medium (ISM) of the host galaxy \citep[e.g.][]{Miyaji2001,Feng2005}.  Indeed, the latter remains a problem for fainter ULXs, as shown by \citet{Earnshaw2016} for a ULX in M51. However, the soft X-ray residuals may also be attributed to processes near the ULX itself \citep[e.g.][]{Roberts2004}, and indeed recent spatial analyses of an object with prominent residuals, NGC 5408 X-1, shows that the residuals are largely constrained to be associated with the ULX itself rather than the surrounding galaxy \citep{Sutton2015}.  

These residuals take a consistent form across several different ULXs with high signal-to-noise {\it XMM-Newton\/} EPIC data \citep{Middleton2015_454}, and can be modeled by absorption from a partly ionized medium, outflowing at $v \approx 0.2c$ \citep{Middleton2014}. These soft residuals have now been resolved in high spectral resolution data. \citet{Pinto2016} see both rest-frame emission and blueshifted absorption in the {\it XMM-Newton\/} RGS spectra of NGC 1313 X-1, and interpret the latter as originating in the optically thin phase of a wind flowing toward us at $0.2c$. They also report similar features with lower significance and an outflow speed of $0.22c$ from NGC 5408 X-1. Further evidence for a mildly relativistic outflow from NGC 1313 X-1 is provided by the detection of a highly ionized Fe K absorption feature that is blueshifted to the same extent as soft X-ray absorption features \citep{Walton2016}.


\section{X-RAY TIMING}
\label{sec:timing}

X-ray timing of Galactic X-ray binaries has lead to the identification of neutron star accretors from coherent pulsations, measurement of orbital and superorbital periods, and the discovery of a rich phenomenology of fast timing behaviors related to the accretion state and mass of the compact objects. While X-ray timing offers great promise for understanding the physical nature of ULXs, a large number of photons and/or extensive monitoring campaigns are needed for timing analysis. High signal-to-noise power spectra and extensive monitoring campaigns have been obtained only for a handful of ULXs. 

\begin{figure}[tb]
\includegraphics[width=3.5in]{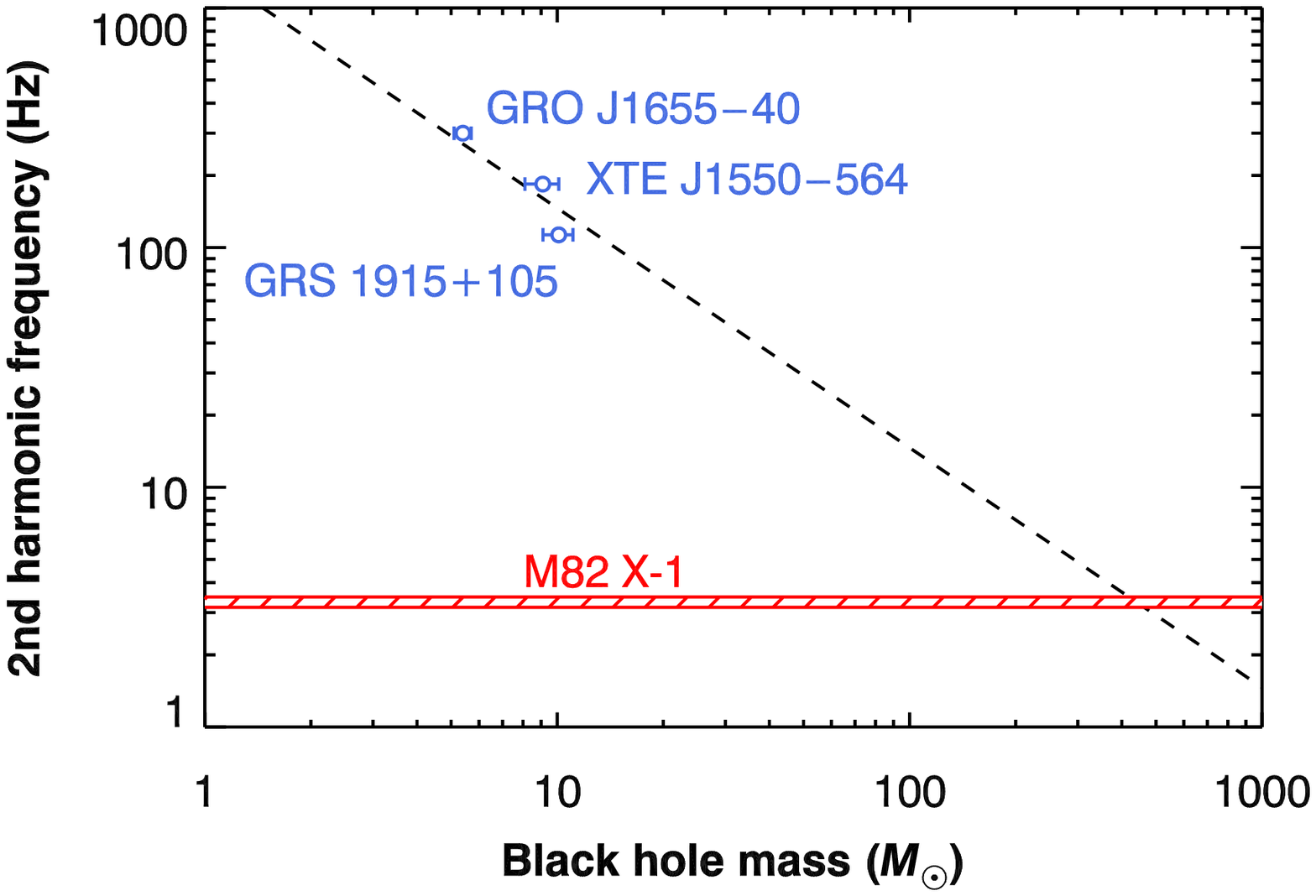}
\caption{High-frequency QPO frequency (second harmonic) versus black hole mass. The systems in order of increasing mass are for GRO J1655$-$40, XTE J1550$-$564 and GRS 1915+105. The dashed line is a linear regression between the mass and the QPO frequency deduced from stellar mass black holes. The hatched region is the measured frequency error interval for M82 X-1 and a black hole mass of $\sim$400~$M_\odot$ can be inferred  (Pasham et al.  2014). All errors are 90\% confidence intervals.}
\label{fig:hfqpo}
\end{figure}

\subsection{Ultraluminous X-ray pulsars}

The most striking finding with fast X-ray timing was the discovery with {\it NuSTAR} of coherent X-ray pulsations from a ULX in M82 \citep{Bachetti2014}. The signal arose from M82 X-2, a transient X-ray source with a peak luminosity above $10^{40} \rm \, erg \, s^{-1}$,  located in the crowded core of the central starburst of M82. The 1.37-second pulsation is superposed on a 2.5-day orbital modulation and shows continuous spin-up at $\dot{P} = -2 \times 10^{-10} \, \rm s \, s^{-1}$. The coherent pulsations are the clear signature of a rotating neutron star, excluding the possibility of a black hole accretor and indicating that the apparent luminosity far exceeds the Eddington limit.


Coherent pulsations have recently been found from two additional ULXs. The three ULX pulsars share similar temporal and spectral properties, indicative of a new type of X-ray sources driven by similar physical processes. NGC 7793 P13 shows pulsations at 0.43~s with a secular derivative $\dot{P} = -3 \times 10^{-11} \, \rm s \, s^{-1}$ over a span of 2.5 years \citep{Israel2016_NGC7793,Fuerst2016}. This source is unique amongst ULX in that the companion star has been identified from absorption lines in the optical spectrum and a dynamical mass constraint is available, see Section~\ref{sec:companion}. NGC 5907 ULX-1 is the most luminous pulsar discovered so far, with a peak luminosity about $10^{41} \rm \, erg \, s^{-1}$, which is $\sim$500 times the Eddington limit of a neutron star \citep{Israel2016_NGC5907}. From 2003 to 2014, its pulsation period changed from 1.43~s to 1.13~s, suggestive of a secular $\dot{P} = -8 \times 10^{-10} \, \rm s \, s^{-1}$.

In all three ULX pulsars, the pulse profile is nearly sinusoidal. The pulsed fraction is relatively low, $\sim 10\%$, below 2~keV, and can increase to as high as 50\% at energies above 8~keV \citep{Israel2016_NGC7793}. The sinusoidal profile suggests that the emission is not strongly beamed. All three ULX pulsars show X-ray spectra typical of those found in the UL state; however, the spectral turnover in the pulsed emission M82 X-2 is quite high at $\sim 14$~keV \citep{Brightman2016}. All the three ULX pulsars are highly variable sources. They exhibit a luminous phase with peak luminosity of $10^{40} - 10^{41} \rm \, erg \, s^{-1}$ and flux variation of $\sim10$ during which pulsations are sometimes detected along with a faint phase with fluxes below $10^{37} - 10^{38} \rm \, erg \, s^{-1}$. This high level of variability, by factors of $\sim 40$ to 1000 or more, is unusual for ULXs. Physical models of the ULX pulsars will be discussed in Section~\ref{sec:nssupereddmodels}.

\subsection{Incoherent Fast timing}

The fast timing properties of GBHBs have been well studied mostly based on observations with RXTE \citep{Remillard2006}. The orbital frequency at the inner edge of the accretion disk around a non-rotating $20 M_{\odot}$ black hole is 110~Hz and scales inversely with mass. This provides a natural division between high-frequency quasi-periodic oscillations (HF-QPOs) that are thought to be produced by dynamics near the inner edge of the accretion disk and the timing noise at lower frequencies which usually consists of a broad-band continuum and sometimes one or more low-frequency QPOs that may be related to dynamics farther out in the disk or in the corona.

\subsubsection{High-frequency variability}

HF-QPOs in GBHBs occur at fixed frequencies related by harmonic ratios, most often 3:2 \citep{Remillard2006}. They have relatively high coherence, $Q = $ centroid frequency/width, in the range 5 to 30. The HF-QPOs are strongly detected mostly in the steep power-law state and sometimes in the thermal state, but their frequencies are independent of the source luminosity and are found to be inversely scaled with the black hole mass. This has been interpreted as evidence that the HF-QPO frequency is determined solely by the black hole mass and spin. \citet{Abramowicz2004} suggested that if ULXs harbor IMBHs, then they should produce HF-QPOs at about 1~Hz and that detection of two QPOs in a 3:2 frequency ratio would be strong evidence in favor of an HF-QPO interpretation.

By stacking RXTE observations of M82 in a time span of six years, \citet{Pasham2014} discovered a pair of QPOs centered at $3.32 \pm 0.06$ and $5.07 \pm 0.06$ Hz, in a harmonic ratio of 3:2 similar to the high-frequency QPOs seen in GBHBs. The QPOs have high coherence, $Q > 27$, and can be detected only via the addition of large numbers of RXTE observations of M82 which suggests that they have stable centroid frequencies. The significance of the individual QPOs detections is low, $3.7\sigma$ and $2.7\sigma$, respectively, while the joint significance is $4.7\sigma$. Scaling from GBHB, see Fig~\ref{fig:hfqpo}, leads to a mass estimate of $428 \pm 105 M_\odot$ for the compact object mass in M82 X-1 \citep{Pasham2014}. The high coherence, accurate 3:2 frequency ratio, and frequency stability of these QPOs is strong evidence in favor of interpreting them as analogous to the HF-QPOs seen from GBHBs. 

\citet{Pasham2015} also claimed detection of a harmonic QPO pair, centered at 0.29 and 0.46~Hz, in NGC 1313 X-1 with {\it XMM-Newton} data. However, the 0.29~Hz QPO has low coherence, with $Q = 2.2$, and the 0.46~Hz QPO is not highly significant. Furthermore, the inferred black hole mass would be $\sim 5000 M_\odot$ giving a luminosity of $0.03 L_{\rm Edd}$ that is inconsistent with the fractional Eddington rate at which HF-QPOs are detected in GBHBs. 

The detection of HF-QPOs from ULXs offers a potential means to determine the compact object masses. The results from M82 X-1 are consistent with spectral results suggesting a mass in the range of a few hundred solar masses (Section~\ref{sec:m82x1}). Confirmation of that result using a future X-ray observatory with larger collecting area and searches for HF-QPOs from other ULXs would be of great interest.

\subsubsection{Low-frequency variability}
\label{sec:lowfreqvar}

In GBHBs it has long been established that the accretion states are distinctive in terms of both X-ray spectral and timing properties \citep[e.g.][]{Remillard2006}. Indeed, they are defined by a combination of both. It is now beginning to emerge that ULXs may show similar links between their spectra and timing characteristics.

\medskip\noindent {\bf Rapid variability.} The variability of a source can be characterized by binning the data in time and calculating the excess variance over that expected from Poisson fluctuations. This is often quantified as the `fractional variability', $f_{\rm var}$. One can also calculate the power spectral density and integrate the excess power above Poisson fluctuations over a fixed frequency interval. This is quantified as the root-mean-squared (rms) power, sometimes normalized by the average flux (fractional rms) \citep{Vaughan2003}. 

\citet{Heil2009} conducted a timing survey of 16 bright ULXs and found that 6 of them showed significant intrinsic time variation, while, in contrast, 3 sources displayed suppressed variation compared with bright GBHBs and AGNs in the frequency range from 1 mHz to 1 Hz. \citet{Sutton2013_435} compared $f_{\rm var}$ (with 200~s bins) to energy spectra measured in the same observation and found that hard UL spectra, with 1--10~keV flux larger than 0.3--1~keV flux, do not show $f_{\rm var} > 10\%$ in the $0.3-10$ keV band. In contrast, some softer UL spectra and a minority of BD spectra show much higher $f_{\rm var}$, between 10--40\%, but this variability appears transient, with objects showing high $f_{\rm var}$ in some observations but not others. \citet{Middleton2015_447} built on this by calculating the rms amplitude in the $3-200$~mHz band. Their results show a similar trend, with some but not all soft UL spectra showing high $f_{\rm var}$. They also show that in some instances hard UL spectra can have high $f_{\rm var}$, but that it tends to manifest at their lowest observed luminosities.

\citet{Sutton2013_435} showed that the variability is more pronounced in the 1--10~keV band than in the 0.3--1~keV band. This simple comparison can be improved upon by calculating the covariance spectra which selects out variations correlated with those in a reference energy band, chosen to have strongly significant variability, and removes uncorrelated noise, such as Poisson fluctuations. \citet{Middleton2015_447} derived covariance spectra for several ULXs and found that the shape of the variable component closely matches the shape of the hard component in the full ULX spectrum, implying that the hard component produces the variability of the ULXs on short timescales.

SSULXs show strong variability with fractional rms amplitudes reaching $\sim$50\% or higher \citep{Soria2016,Jin2011}. The variability increases for higher energy bands, but is still quite strong in the lowest energy bands accessible, 0.3--0.7~keV, implying that there must be strong variability in the soft component which dominates the emission.  Thus, the timing properties of SSULXs appear to be distinct from those of broadband ULXs \citep{Feng2016}.

\medskip\noindent {\bf The rms-flux relation.} \citet{Uttley2001} found that there is a simple linear scaling of rms variability amplitude and flux for many GBHBs and AGN: the variability is higher when the flux is higher. \citet{Heil2010} measured a positive rms-flux relation in the 1-10~keV band for NGC 5408 X-1 on time scales shorter than days, over which the timing properties are stationary. \citet{Hernandez-Garcia2015} detected a similar linear rms-flux relation for NGC 6946 X-1 and in additional observations of NGC 5408 X-1. They were unable to constrain the relation in other ULXs due to a lack of either statistics or source variability. The results suggest that a common physical mechanism produces the fast variability in ULXs, GBHBs, and AGN. Interestingly, \citet{Caballero-Garcia2013_435} found that for NGC 5408 X-1 the rms is anti-correlated with the 1-10~keV flux on timescales of years. Similar behavior is found in the bright hard state of GX 339-4, but differences in the energy spectra argue against a common accretion state.

\medskip\noindent {\bf QPOs and broadband noise.} Beyond the total variability, the power spectra of accreting objects reveal a complex, and unfortunately poorly understood, array of timing features that include continuum noise, with breaks between different components, and QPOs, \citep[see Fig.~2 of][]{Remillard2006}. The characteristic frequencies vary with the accretion rate or spectral properties as well as with the black hole mass \citep{McHardy2006}.

Low-frequency QPO and timing features such as the low-frequency break in the continuum noise have been reported only for a small number of ULXs due to the large photon statistics needed and the fact that many ULXs have little fast variability. The first QPO detection was from M82 \citep{Strohmayer2003} later localized to M82 X-1 \citep{Feng2007}. Additional observations revealed shifts in the QPO frequency in the range 34--120~mHz and a flat-top continuum component with a low frequency break around 34~mHz \citep{Dewangan2006,Mucciarelli2006,Caballero-Garcia2013_436}. The low frequency noise and QPOs disappeared in three deep {\it XMM-Newton} observations, suggesting a change in the accretion state of M82 X-1 \citep[][see Section~\ref{sec:imbh}]{Feng2010}. NGC 5408 X-1 shows strong timing noise \citep{Soria2004}, a low frequency break at a few mHz, and QPOs with varying frequency in the 10--20~mHz range \citep{Strohmayer2009}.  Low frequency QPOs around 33 and 80~mHz, above the break of a flat-topped noise component, were detected in NGC 1313 X-1 \citep{Pasham2015}. \citet{Rao2010} reported QPOs at 8.5~mHz above a continuum break near 3~mHz in NGC 6946 X-1. An integrated variation amplitude of 60\% in the frequency range of 1-100~mHz makes it one of the most variable ULXs.

Whether or not the black hole mass can be robustly determined from low-frequency QPOs detected from ULXs is controversial. In GBHBs, low frequency QPOs can be divided into three types, A, B, and C \citep{Casella2005}. Most low-frequency QPOs seen in GBHBs are of type-C, which are strong and associated with a band-limited noise continuum \citep{Vignarca2003}. The QPO frequency varies in tight correlation with the energy spectral index and the disk flux, but the index saturates at high frequency. The black hole mass can be estimated via QPOs only if they are type-C and in the unsaturated phase of the correlation. Simply based on the QPO amplitude and coherence, and the association with a flat-topped noise continuum, the QPOs detected in M82 X-1 and NGC 5408 X-1 are similar to those of type-C. \citet{Pasham2013} examined all of the {\it XMM-Newton} observations of M82 X-1 and found that the QPO centroid frequency was correlated with the source count rate, but uncorrelated with the hardness ratio, an indicator of the spectral index. This challenges the use of these QPOs to measure the black hole mass. For NGC 5408 X-1, \citet{Middleton2011_411} questioned the identification of type-C for the QPOs on the basis that the QPO frequencies were not correlated with the  continuum noise break frequencies. \citet{Dheeraj2012} found that the QPO frequency was independent of the spectral parameters. Again, the QPOs are either not of type-C or in the saturated phase of the correlation, calling into question their use in estimating the black hole mass.

\subsection{Long-term periodicities and quasi-periodicities}

The detection of orbital periodicity is of great interest. For Roche-lobe overflow systems, the orbital period directly determines the average companion star density and consequently constrains its spectral type. Also, knowledge of the orbital period is essential to measuring the optical mass function and dynamically measuring the compact object mass. 

RXTE monitoring of M82 showed a 62-day period \citep{Kaaret2006}. The amplitude of the periodic flux variation suggests that M82 X-1 is its source; the sinusoidal variations are incompatible with an origin from M82 X-2 which shows a bi-modal flux distribution \citep{Tsygankov2016}. The high coherence, $Q = 22.3$ over the data span of 1124 days (18 cycles), suggests that the modulation is due to orbital motion \citep{Kaaret2007}. However, continued monitoring revealed that the modulation phase changed by 0.4 after a major outburst and was interpreted as evidence in favor of a superorbital origin \citep{Pasham2013_774}. However, the phase of the X-ray minima of the 13 day orbital modulation of SS 433 is known to depend on the 162 day precession period of the jets \citep{Atapin2016}, hence the orbital interpretation is not excluded.

A long-term periodic modulation around 115~days was measured from NGC 5408 X-1 with Swift observations \citep{Strohmayer2009}. The modulation was later found to have disappeared after a few cycles \citep{Grise2013} and a wavelet analysis showed that the period shifted between 115 and 136 days \citep{An2016} suggesting that the modulation is superorbital \citep{Foster2010}. Swift monitoring detected a long-term periodicity at $\sim$78 days from NGC 5907 ULX-1 \citep{Walton2016}. The periodicity was detected over a course of $\sim$700 days or nearly 10 cycles suggesting that the periodic variation was stable and due to orbital modulation, but future observations are needed to test this scenario.

Recently, \citet{Urquhart2016} reported on the discovery of the first two ULXs with X-ray eclipses, both in M51 and with a peak luminosity of about $2 \times 10^{39} \rm \, erg \, s^{-1}$. This implies that the two systems are viewed close to edge-on. Future observations are needed to determine the orbital period. If an orbital modulation is confirmed, they may be good candidates for dynamical mass measurements. 

Repeated outbursts, each showing a fast rise and exponential decay, have been seen in the lightcurve of the IMBH candidate ESO 243-49 HLX-1 with the recurrence time increasing from less than 400 days to nearly 500 days over six cycles \citep{Lasota2011,Servillat2011,Yan2015}. Models proposed to explain the behavior include: orbital evolution of a star captured in a highly eccentric orbit star around an IMBH due to tidal effects and mass transfer \citep{Godet2014} but this is controversial \citep{vanderHelm2016}; disk instabilities causing super-orbital modulation in a super-Eddington system \citep{Lasota2015}; precession of an X-ray beam \citep{King2016_458}; and an IMBH fed by winds from a giant star with a tidally stripped envelope \citep{Miller2014}. 


\section{MULTI-WAVELENGTH COUNTERPARTS}
\label{sec:multi}

Optical (including infrared and ultraviolet) and radio observations of ULXs provide important information beyond what can be gleaned from X-rays. The identification of point-like optical counterparts could permit measurement of radial velocity curves for the companion stars and thus unambiguous constraints on the compact object masses. However, in many cases the point-like optical emission is dominated by light reprocessed by accreting matter. This outshines the companion star, but provides information about the accretion flow. Many bright ULXs are found to be spatially associated with an optical \citep{Pakull2003} and/or radio nebula \citep{Kaaret2004} powered by radiation or matter outflows. Finally, optical study of the environments of ULXs can shed light on the evolutionary history of the binaries. 

\begin{figure}[tb]
\includegraphics[width=3in]{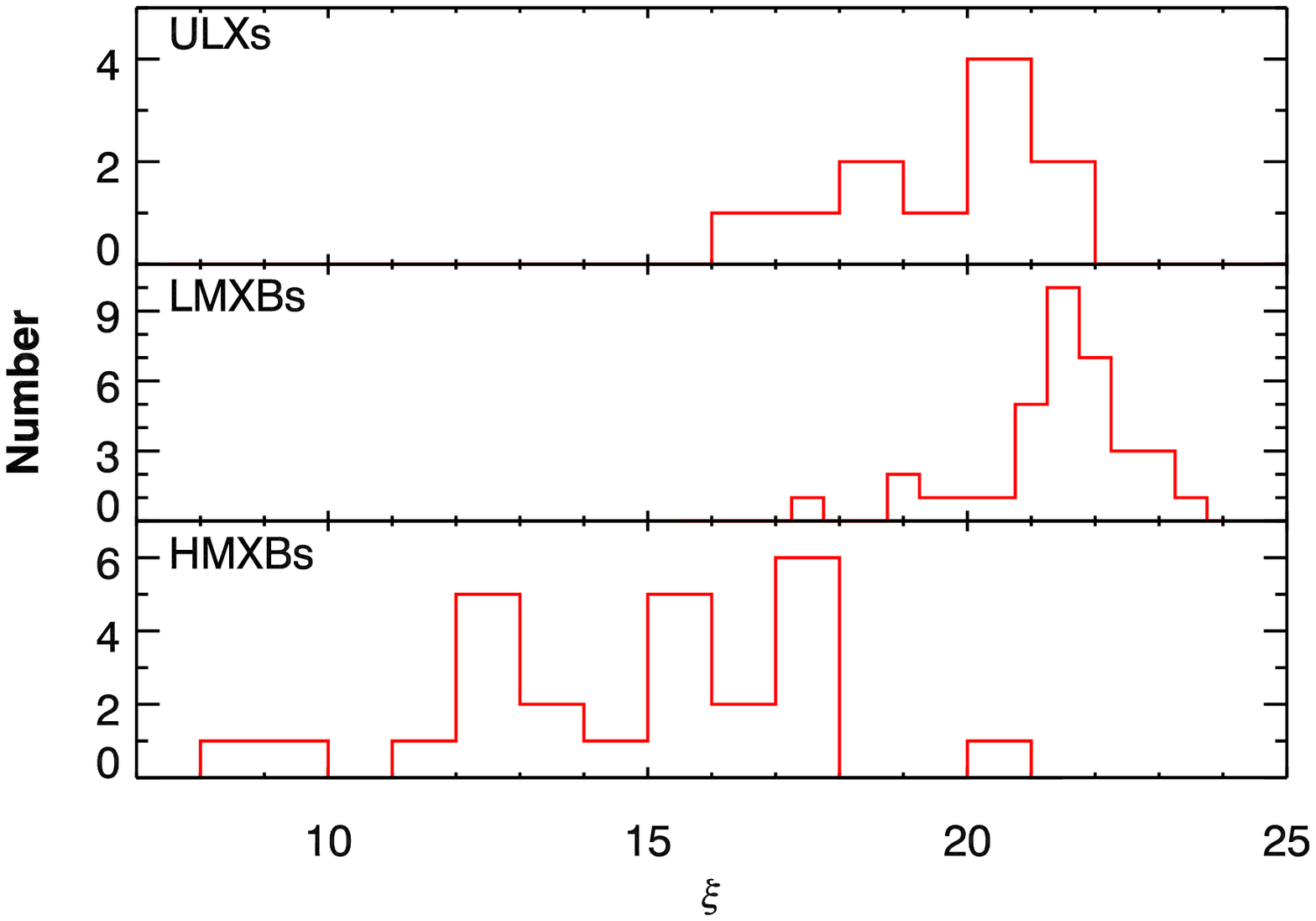}
\caption{X-ray to optical flux ratio of ULXs with unique point-like counterparts found in HST images from \citet{Tao2011}. The ULX flux ratios are similar to those of active LMXBs suggesting that the optical emission arises from the accretion flow. The figure does not include supersoft ULXs.}
\label{fig:xrayopt}
\end{figure}

\subsection{Optical Photometry}

The point-like optical counterparts of ULXs are dim, typically with optical magnitudes $\ga 21$. Most ULX candidates with brighter optical counterparts are identified as background AGN or foreground stars. ULXs tend to lie in crowded fields, hence identification of unique counterparts often necessitates use of the {\it Hubble Space Telescope} (HST) for optical imaging, {\it Chandra} for X-ray source localization, and specialized analysis to improve the relative optical/X-ray astronomy to the $0.2$ arcsecond level. Unique point-like optical counterparts have been identified for $\sim 20$ ULXs \citep{Tao2011,Gladstone2013}. The V-band absolute magnitudes are bright, in the range $M_V = -3 {\rm \, to} -8$, and the colors tend to be blue, with $B - V$ in the range $-0.6$ to $+0.4$. This would suggest OB supergiant companions; however, when multiple band photometry is available it is usually not consistent with any spectral classification.

The X-ray to optical flux ratio, defined as per \citet{vanParadijs1995} as $\xi =  B_0 + 2.5 \log F_X$ where $B_0$ is the dereddened $B$ magnitude and $F_X$ is the observed 2-10 keV flux in $\mu$Jy, of most ULXs is very high and similar to that of active LMXBs rather than HMXBs, see Figure~\ref{fig:xrayopt}. The optical emission from X-ray bright LMXBs is dominated by X-ray heating of the accretion disk and/or companion star. The similarity in flux ratios suggests that optical emission from ULXs is likely dominated by light from the accretion flow. The few ULXs for which repeated HST observations are available show variability in both magnitude and color, again suggesting origin of the optical light from reprocessing rather than the intrinsic emission of the companion star \citep{Tao2011}.

\begin{figure}[tb]
\includegraphics[width=2.0in]{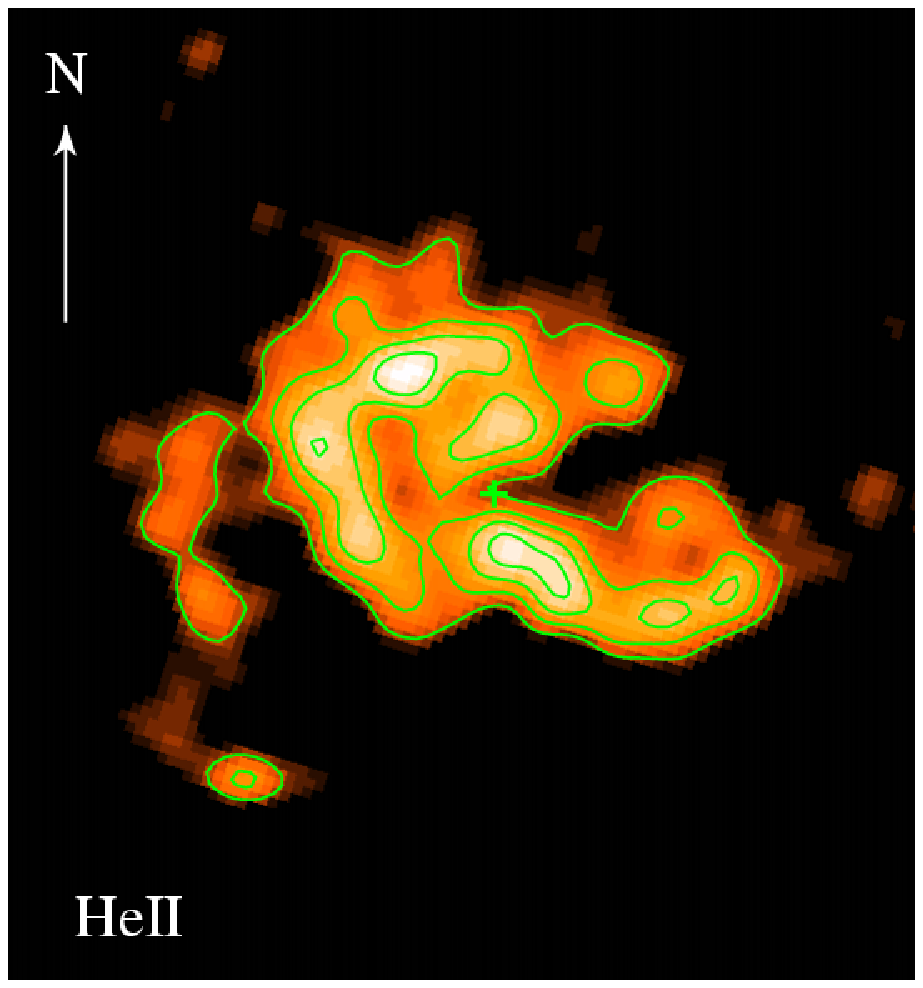}
\caption{HST image of He~{\sc ii} $\lambda$4686 line emission surrounding Holmberg II X-1 from \citet{Kaaret2004}. The ULX position is marked with a green cross. The arrow points North and has a length of $1$ arcsecond (15~pc); East is to the left. The green curves are contours of the He{\sc ii} emission with levels of $[2,3,4,5] \times 1.9 \times 10^{-16} \rm \, erg \, cm^{-2} \, s^{-1} \, arcsec^{-2}$.}
\label{fig:hoii_heii}
\end{figure}

\subsection{Nebular Counterparts}
\label{sec:nebulae}

Some ULXs for which sensitive optical imaging is available are coincident with an optical nebula. These nebulae tend to be large with extents of tens to hundreds of parsecs. Most appear to be powered by shocks created by outflows from the binary interacting with the surrounding medium, while, in a few cases, the nebulae are powered by photoionization \citep{Pakull2003}.

Strong high ionization lines, particularly He~{\sc ii} $\lambda$4686 and [Ne~{\sc V]$\lambda$3426}, provide evidence for photoionization. The He~{\sc ii} line is produced only by fully ionized He, which requires photons with energies in excess of 54.4~eV. Thus, the image of He~{\sc ii} line emission surrounding Holmberg II X-1 shown in Figure~\ref{fig:hoii_heii} is essentially a map of the soft X-ray illumination of the nebula by the ULX. The map is direct evidence that the ULX is at most mildy beamed as a highly beamed source would illuminate only a small region around our line of sight to the ULX. Precession of a narrow beam could illuminate a wide field; however, if the X-ray beam is significantly narrower than the precession cone angle, then the X-ray flux should usually be low -- which is not observed in long term monitoring with {\it Swift} \citep{Grise2010}.

The luminosity of the isotropic nebular emission can be used to estimate the true soft X-ray luminosity of the ULX \citep{Pakull2003}. The best constraint comes from the [O {\sc IV}] 25.89 $\mu$m emission line which implies a lower bound on the bolometric luminosity of Holmberg II X-1 of $1.1 \times 10^{40} \rm \, erg \, s^{-1}$ \citep{Berghea2010}. The recombination time of He$^{++}$ in the nebula is roughly 3000~yr, for an electron density of $10 \rm \, cm^{-3}$ and a temperature of 20,000~K, suggesting that the average luminosity has been at least $10^{40} \rm \, erg \, s^{-1}$ for several thousand years.

Prominent lower ionization lines, such as [O~{\sc i}] $\lambda$6300, suggest shock ionization. The shocks in such ULX nebulae are moderately strong with velocities of 20--100~km/s. The outflow powers, which can be estimated from the line luminosities \citep{Abolmasov2007} or kinematics, are comparable to or greater than the X-ray luminosities, sometimes suggesting super-Eddington mechanical power \citep{Soria2014}. An extreme case is a shock-powered X-ray emitting bubble in NGC 7793 with a mechanical power of a few $10^{40} \rm \, erg \, s^{-1}$  that exceeds the X-ray luminosity by a factor of $10^{3}$ \citep{Pakull2010}. The expansion speeds are often supersonic, reaching 150~km/s. The speeds together with the nebular sizes imply characteristic ages on the order of $10^6$~yr. The total nebular energy is of the order $10^{52}$~erg. This is an order of magnitude larger than the typical total mechanical energies of standard supernovae \citep{Roberts2003}. 

Radio nebulae are also observed around a few ULXs \citep{Kaaret2003}. The spectra are optically thin suggesting synchrotron emission from mildly relativistic electrons accelerated in the shocks produced by the interaction of an outflow from the ULX with the surrounding interstellar medium. The minimum total energy needed to power the radio emission can be estimated via equipartition between the energy of relativistic particles and the magnetic field. Typical values are around $10^{49}$~erg to $10^{50}$~erg, although this depends on the assumed electron spectrum \citep{Miller2005,Lang2007}. For IC 342 X-1, the minimum energy required to power the radio nebula is $9 \times 10^{50}$~erg, while the total energy of the shock powered nebula estimated from optically measured expansion and size is $5 \times 10^{52}$~erg \citep{Cseh2012}. Thus, about 2\% of the total energy is in relativistic particles.

Shock-ionized ULX nebulae are powered by outflows. The outflows could be sub-relativistic winds, continuous relativistic jets, or episodic ejections of relativistic particles. There is observational evidence for powerful outflows in X-ray spectroscopy of ULXs (Section~\ref{sec:xrayoutflows}) and such outflows are also predicted by the theory of super-Eddington accretion (Section~\ref{sec:superedd}). AGN and GBHBs at high accretion rates tend to show strong winds rather than relativistic jets, so this is consistent with the general picture of most ULXs as super-Eddington systems. An interesting counterexample is Holmberg II X-1 that has a photoionized optical nebula and a triple-lobed radio nebula in which the central components evolve on time scales of a few years \citep{Cseh2015}. Its radio emission is likely powered by episodic ejections of relativistic particles, similar to those seen from GBHBs at state transitions in major outbursts.

\subsection{Compact radio counterparts}
\label{sec:compactradio}

In AGN and GBHBs, compact radio jets are detected in the X-ray hard state that predominately occurs at low accretion rates. In such states, there is a relationship between X-ray luminosity $L_{\rm X}$, radio luminosity $L_{\rm R}$, and black hole mass $M$ such that $L_{\rm X} = \xi_R \log L_{\rm R} + \xi_M \log M +b_X$ \citep{Merloni2003}. The coefficients ($\xi_R, \xi_M, b_X$) are fixed by fitting to observations of large sets of objects, including both GBHBs with stellar-mass black holes and AGN with supermassive black holes. Thus, the relation interpolates across the IMBH regime rather than extrapolating from only GBHBs or AGN. Use of this relationship to constrain the compact object masses in ULXs has generated great interest.

\citet{Mezcua2015} report detection of a persistent, compact radio jet from the ULX NGC 2276-3c using the European Very Long Baseline Interferometry Network (EVN). The source exhibits a peak X-ray luminosity of $6 \times 10^{40}$~erg/s and has a hard X-ray spectrum, suggestive of the GBHB hard state. The EVN revealed a 1.8~pc radio jet oriented along its large, 650~pc, radio lobes. The jet kinetic power is close to its X-ray luminosity. Using the fundamental plane relation, its mass is estimated to be $5 \times 10^{4} ~ M_{\odot}$ with an uncertainty of 0.7 dex.  The location of this IMBH candidate in a spiral arm with unusual morphology and a high star formation rate suggest that it may be the nucleus of a stripped dwarf galaxy \citep{Mezcua2015}. However, \citet{Yang2016} reanalyzed the same data and failed to confirm the detection.

Flaring radio emission has been reported from ESO 243-49 HLX-1 \citep{Webb2012} and XMMU J004243.6+412519 in M31 \citep{Middleton2013}. The M31 source produced a thermal-dominant spectrum in the same outburst, suggesting that the source was not in the X-ray hard state at the epoch of radio emission and that the radio emission was likely due to discrete ejecta similar to that seen from X-ray soft, radio flaring GBHBs such as GRS 1915+105. The peak X-ray luminosity of $1.3 \times 10^{39}$~erg/s of the M31 source is also similar to that of GRS 1915+105, suggesting the source contains a stellar-mass black hole radiating close to its Eddington limit.  \citet{Webb2012} detected radio flares from  ESO 243-49 HLX-1 after the source reached its peak luminosity, while the source was already in a disk-dominated state. Thus, the fundamental plane relation does not apply. They derived an upper limit on the black hole mass of $9 \times 10^4 M_{\odot}$ by assuming that the radio flares occurred when the X-ray luminosity is $(0.1 - 1) L_{\rm Edd}$; however, flares are known to occur from AGN and SGR A* at much lower X-ray luminosities. \citet{Cseh2015} reported detection of radio emission associated with a hard X-ray spectrum, but application of the fundamental plane is not very constraining as $M < 3 \times 10^{6} M_{\odot}$.

\begin{figure}[tb]
\includegraphics[width=2.5in]{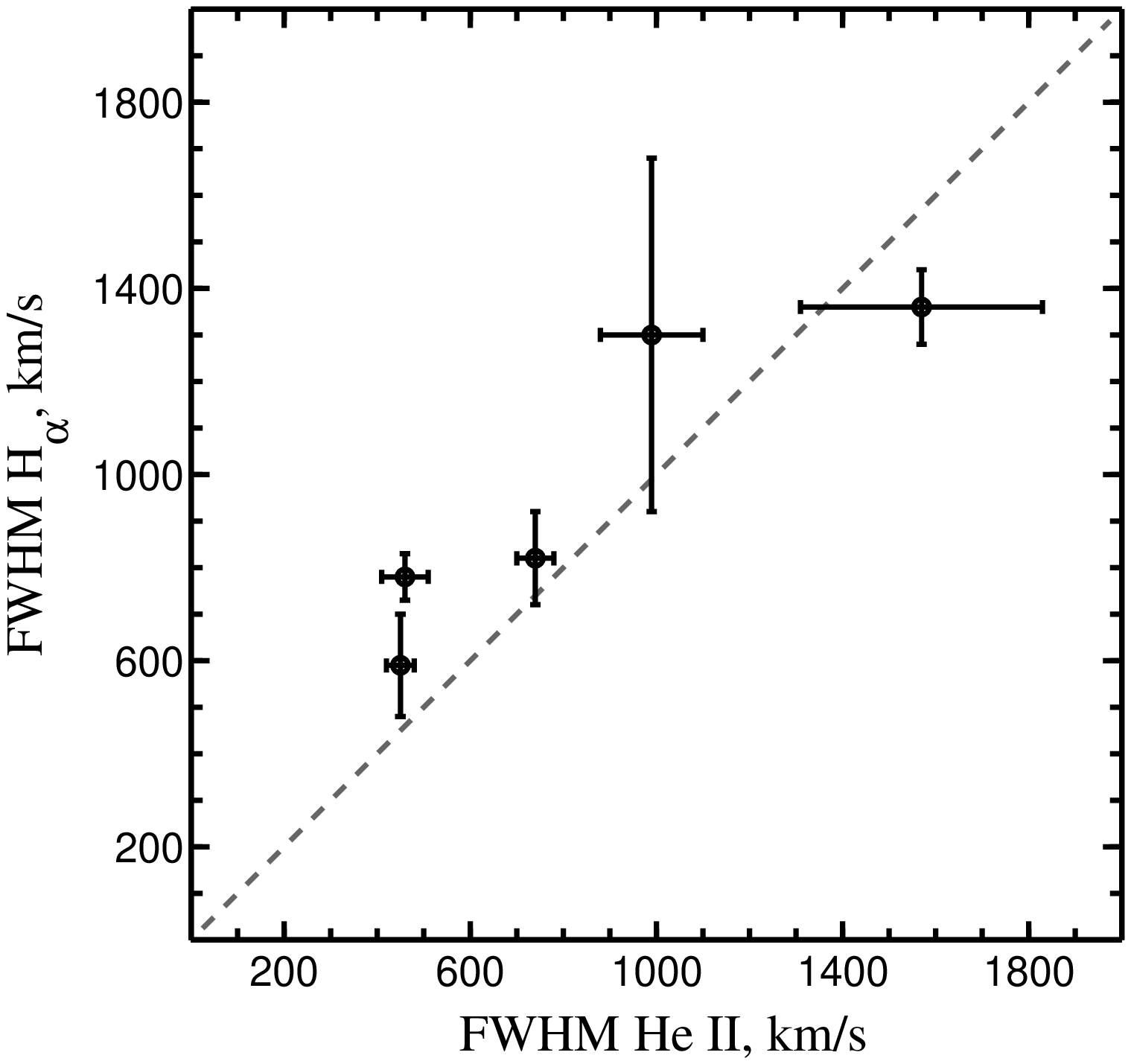}
\caption{Emission line widths of He{\sc ii} and H$\alpha$ in the spectra of ULX optical counterparts, from \citet{Fabrika2015}. From left to right, the points are for Holmberg II X-1, NGC 4559 X-7, NGC 5408 X-1, Holmberg IX X-1, and NGC 5204 X-1.}
\label{fig:heii_width}
\end{figure}

\subsection{Optical evidence for outflows}
\label{sec:optoutflows}

Optical spectra of ULXs also reveal broad emission lines with widths of 500-1500~km/s, notably in the Balmer series and in He~{\sc ii} $\lambda$4686. The lines can be singly or doubly peaked with radial velocities of up to $800$~km/s \citep{Roberts2011,Cseh2013, Fabrika2015}. Such lines could be produced via irradiation of an accretion disk or in a wind. In a disk, He~{\sc ii} lines are produced in regions of higher excitation than the Balmer lines. These regions are closer to the compact object, hence the He~{\sc ii} lines should be significantly broader and are often double peaked \citep[e.g.][]{Soria2000}. Figure~\ref{fig:heii_width} shows that the opposite is true in ULXs. \citet{Fabrika2015} suggest that the lines are produced in a radiatively accelerated wind. The more highly ionized gas is still located closer to the compact object, but has lower velocity since the wind is accelerated as it flows out. This origin of the optical emission would preclude attempts to use the emission lines to measure radial velocity curves or to use irradiated disk models to constrain the parameters of the binary system.

The optical spectra of the SSULX M81 ULS-1 show a remarkably blue shifted H$\alpha$ line \citep{Liu2015}. The line appears in multiple observations at varying blue shifts corresponding to projected velocities of 0.14$c$--0.17$c$ indicating that it originates in a relativistic baryonic jet. The X-ray spectra of M81 ULS-1 are well fitted with a cool DBB model with temperatures ranging from 0.07--0.10~keV clearly establishing it as a SSULX. 

These outflows are similar to those seen from SS 433, an X-ray binary within the Milky Way thought to contain a stellar-mass black hole in a 13.1 day orbit, accreting at a super-Eddington rate of $\sim 10^{-4} M_{\odot} \rm \, yr^{-1}$ via Roche lobe overflow from a massive donor star that is likely an evolved A supergiant \citep[for a comprehensive review see][]{Fabrika2004}. SS 433 has precessing relativistic jets revealed by Balmer and He{\sc I} emission lines with strongly varying Doppler shifts reaching 50,000 $\rm km \, s^{-1}$. A kinematic model of the radial velocity curves shows the lines are produced in two opposing baryonic jets moving at $0.26c$ and precessing in a cone with half-angle of 20$^{\circ}$ around an axis inclined by 79$^{\circ}$ from the line of sight. SS 433 also exhibits broad (FWHM $\sim 1000 \rm \, km \, s^{-1}$) He~{\sc ii} lines with Doppler shifts that vary by $\sim 400 \rm \, km \, s^{-1}$. This line is thought to be produced in a slow disk wind generated close to the compact object. Furthermore, SS 433 produces compact radio jets and is surrounded by a radio nebula (W50) with an extent of about 100~pc along the radio/optical jet axis. The similarity of the optical and radio counterparts of ULXs with the corresponding aspects of SS 433 is striking and supports interpretation of ULXs as super-Eddington accretors (Section~\ref{sec:superedd}). 

\subsection{Companion stars and mass constraints}
\label{sec:companion}

The ULX NGC 7793 P13 is unique in that its optical spectrum shows absorption lines that clearly indicate we detect light from the companion star and enable its classification as a B9Ia supergiant \citep{Motch2014} with a mass of 18--23~$M_{\odot}$. The radial velocity curve shows significant variations from year to year that prevent estimation of a mass function. However, periodicity at 63 days is present in both the He~{\sc ii} radial velocity curve and V and u band photometry. Modeling of the photometry implies a compact object mass less than 15 $M_{\odot}$ as larger masses would produce a Roche lobe at periastron smaller than the B9Ia star. As discussed in Section~\ref{sec:timing}, the compact object has been identified as a neutron star based on the detection of coherent pulsations.


The companion star of M101 ULX-1 is classified as a Wolf-Rayet star of WN8 sub-type with a mass of $19 \pm 1 M_{\odot}$ based on the presence of broad He and nitrogen emission lines and the absence of hydrogen and carbon emission lines \citep{Liu2013}. The He~{\sc ii} $\lambda$4686 radial velocity curve suggests a periodicity at 8.2 days and a mass function of $0.18 \pm 0.03 M_{\odot}$. However, the data are sparse and the He~{\sc ii} $\lambda$4686 line could be contaminated by emission from the accretion disk, calling for additional spectroscopy and confirmation of the period via photometry. The inferred orbital parameters imply a minimum compact object mass of $4.4 M_{\odot}$ and allow masses of at least 300~$M_{\odot}$, but \citet{Liu2013} argue that the mass is likely much lower, $\sim$30~$M_{\odot}$, because of the low probability of observing a near pole-on binary assuming a uniform inclination distribution. To power the observed X-ray luminosity via wind-fed accretion, as proposed, the black hole mass must be above 48~$M_{\odot}$.

M101 ULX-1 shows a high level of X-ray variability and the optical spectroscopy was carried out during an expected, although not confirmed, X-ray low state. Optical studies of GBHBs are preferentially carried out during X-ray low or quiescent states to avoid contamination from an irradiated accretion disk. Since most ULXs are consistently X-ray bright, it may be difficult to detect the companion stars and obtain radial velocity curves. 


\subsection{Stellar environments}
\label{sec:environments}

Beyond examining the counterpart of the ULX, HST imaging can also be used to study the stellar environment of the ULX. ULXs in spiral and moderately star-forming dwarf galaxies are sometimes associated with loose clusters or OB associations with masses of a few $10^3 M_{\odot}$ and ages of 10--20~Myr \citep{Feng2008,Grise2008,Grise2011}, but often lie in faint or non-star-forming regions. \citet{Swartz2011} interpret this as evidence that ULXs turn on after the typical lifespan, 10--20~Myr, of an H{\sc ii} region. In large starburst galaxies, bright ULXs tend to be found near dense, compact star clusters or `super star clusters' (SSCs), suggesting a physical association between the ULXs and SSCs \citep{Kaaret2004_348,Poutanen2013}.

\section{SUPER-EDDINGTON ACCRETION}
\label{sec:superedd}

\subsection{Evidence for super-Eddington accretion}
\label{sec:supereddevidence}

The earliest clear evidence for super-Eddington accretion was the detection of coherent pulsations at 69~ms from A0538-66, an X-ray transient in the Large Magellanic Cloud, indicating that it must host a neutron star \citep{Skinner1982}. A0538-66 reaches $L_{X} \sim 8 \times 10^{38} \rm \, erg \, s^{-1}$, apparently exceeding its Eddington limit by a factor of 4. The most remarkable evidence for super-Eddington accretion comes from the recent detection of ULX pulsars containing accreting neutron stars that apparently exceed the Eddington limit by factors up to $\sim 500$ \citep{Israel2016_NGC5907}. For black hole accretors, GBHBs have long been suspected to exceed the Eddington limit, e.g.\ V4641 Sgr \citep{Revnivtsev2002} and GRS 1915+105 \citep{Done2004}. ULXs reach significantly higher luminosities and if some fraction of ULXs host stellar-mass black holes, then they must also be super-Eddington accretors.

There is substantial evidence for strong outflows from ULXs from the presence of shock-powered optical and radio giant bubble nebulae (Section~\ref{sec:nebulae}), broad optical emission lines (Section~\ref{sec:optoutflows}), and blue-shifted X-ray absorption features (Section~\ref{sec:xrayoutflows}). The kinetic energy required in the outflow is often comparable to or larger than the X-ray luminosity and suggests that a large fraction of the total energy release of the accretion flow goes into powering an outflow. These strong outflows are additional evidence in favor of super-Eddington accretion in ULXs. In this section, we examine the mechanisms whereby the mass transfer for super-Eddington accretion may arise and models of the accretion flow near the compact object.

\subsection{Mass transfer}

The first requirement in exceeding the Eddington limit is that there must be a sufficient transfer of material towards the compact object to power the extreme energy release. We can infer the mass transfer rate from the companion star, $\dot{m}$, from the bolometric luminosity $L_{\rm Bol}$ of an accreting system as

\begin{equation}
\dot{m} = {{L_{\rm Bol}}\over{\eta c^2}}
\end{equation}

\noindent where $\eta$ is the radiative efficiency of the accretion flow. Taking a value of $\eta = 0.1$, appropriate for efficient conversion of accreted mass to radiation with little or no outflow, a mass transfer rate of $\approx 1.8 \times 10^{-7} ~ M_{\odot} \rm \, yr^{-1}$ is required for a bolometric luminosity of $10^{39} \rm ~erg~s^{-1}$. As the Eddington luminosity $L_{\rm Edd} \simeq 1.3 \times 10^{38} M_{\rm co} \rm ~erg~s^{-1}$ for a compact object of mass $M_{\rm co}$ (in solar units) accreting hydrogen, the mass transfer rate to reach the Eddington limit can be given as

\begin{equation}
\dot{m}_{\rm Edd} = 2.3 \times 10^{-8} M_{\rm co} ~ M_{\odot} \rm \, yr^{-1}.
\end{equation}

\noindent Hence, a $10 M_{\odot}$ black hole requires a mean accretion rate in excess of $2.3 \times 10^{-7} ~ M_{\odot} \rm \, yr^{-1}$ to exceed its Eddington limit.  This is far in excess of the mass transfer rates in typical Galactic systems -- for example, an LMXB undergoing Roche lobe overflow with an average outburst luminosity $10^{38} \rm ~erg~s^{-1}$, and a duty cycle of 0.01, requires an average mass transfer rate of $\approx 2 \times 10^{-10} ~ M_{\odot} \rm \, yr^{-1}$; a persistent wind-fed HMXB with an average luminosity of $10^{37} \rm ~erg~s^{-1}$ requires $\approx 2 \times 10^{-9} ~ M_{\odot} \rm \, yr^{-1}$ of mass transfer.

How then is such a high mass transfer rate achieved?  The best candidate appears to be thermal-timescale mass transfer in high mass binaries \citep{King2001}, which occurs when the massive secondary star in a binary containing a compact object evolves into the Hertzsprung Gap and expands, so filling its Roche lobe. \citet*{Rappaport2005} used binary evolution models to show that the mass transfer might be high enough during main sequence evolution to fuel a ULX, if the secondary is more massive than $\sim 10 M_{\odot}$; however the accretion rates peaked at a factor $10^2$ higher during the thermal-timescale phase.  The mass transfer rates during this epoch can be very extreme, for example \citet{Wiktorowicz2015} show that a stellar-mass black hole accreting from a similarly-massive star in the thermal-timescale phase can have a mass transfer rate of $10^{-3} ~ M_{\odot} \rm \, yr^{-1}$ for $\sim 10^4$ yr.  They also show a neutron star undergoing thermal-timescale mass transfer from a Helium-burning secondary with mass $\sim$1--2~$M_{\odot}$ will see even higher mass transfer rates of $10^{-2} ~ M_{\odot} \rm \, yr^{-1}$ for $\sim 10^2$~yr.  These results were expanded on by \citet{Pavlovskii2016}, who show that stable mass transfer is possible for a wide range of binary parameters, including for high secondary star/black hole mass ratios, with mass transfer rates of $\sim 1000 \, \dot{m}_{\rm Edd}$ regularly reached during the $10^5$~yr thermal-timescale mass transfer stage. However, this may not be the only situation that can fuel super-Eddington accretion; \citet{King2002} suggests that long-lasting transient outbursts in wide orbital separation LMXBs (such as GRS 1915+105) are capable of depositing material onto a black hole at highly super-Eddington rates. It is clear, then, that the necessary mass accretion rates for super-Eddington ULXs are physically plausible.

\subsection{Neutron-Star Accretion}
\label{sec:nssupereddmodels}

Neutron star accretion is a complex situation since the stellar radiating surface and potentially high magnetic field may strongly influence the accretion process. The discovery of pulsations in M82 X-2 (Section~\ref{sec:timing}) has prompted the development of models to explain how to apparently exceed the Eddington limit by a factor $\ga 100$. 

A key parameter is the strength of the neutron star magnetic field, while two key observational diagnostics on the accretion rate are the luminosity and the pulse frequency derivative.  It is usually assumed that the flow at large radii will be via a standard accretion disk.  At the Alfv\'en or magnetospheric radius, $r_m$, defined as the location where the magnetic pressure balances the ram pressure of the flow, the accretion flow transitions from following Keplerian orbits to following magnetic field lines. However, accretion onto the neutron star surface will proceed only when the angular velocity of the accretion disk at $r_m$ is higher than that of the neutron star; so that the gravitation force at the boundary exceeds the centripetal acceleration of matter tied to the magnetic field lines.  If the accretion rate is low, the magnetospheric radius recedes from the star and accretion is stopped or greatly slowed.  This is the `propeller effect'. The magnitude of $r_m$ and the difference between the Keplerian angular velocity at $r_m$ and the angular velocity of the neutron star determine the torque produced by the accretion flow on the neutron star and hence the secular pulse frequency derivative (the frequency also changes due to orbital motion).

For strong fields, the accretion flow is magnetically threaded on to the neutron star and near the neutron star surface, the flow has the geometry of a column or funnel. This provides a geometric means to exceed the Eddington limit as radiation can escape from the sides of the funnel, perpendicular to the flow. \citet{Basko1976} estimate a maximum enhancement of $\sim 6$.  If the magnetic field is sufficiently high, $B > 10^{13}$~G, near the funnel, then the field will restrict the motion perpendicular to the field lines of electrons in the flow, reducing their scattering cross-section to below the Thomson value for photons polarized perpendicular to the magnetic field. This provides another means to exceed the Eddington limit.

\citet{Bachetti2014} estimate the accretion rate onto M82 X-2 via the pulse frequency derivative assuming that the system is in spin equilibrium as may be expected given the short spin-up timescale, $P/\dot{P} \approx 300$~yr.  They find a rate only a few times higher than Eddington and infer $B \ga 10^{12}$~G.  They suggest that the large apparent luminosity is due to beaming.  However, the sinusoidal pulse profile is inconsistent with the narrow beam expected if there is significant beaming, unless we view a fan beam at a favorable inclination.  This appears untenable now that three ULX pulsars all show sinusoidal pulse profiles.

\citet{Eksi2015} suggest that the mass accretion rate was underestimated because the torque is due to differential rotation between the disk and neutron star at $r_m$ and is therefore small if the system is near spin equilibrium.  They estimate that the dipole field is $B = 7 \times 10^{13}$~G and suggest that there may be stronger multipole fields near the neutron surface that could allow the observed luminosity via opacity reduction. \citet{Mushtukov2015} similarly suggest that luminosities of $\sim 10^{40} \rm \, erg \, s^{-1}$ can be achieved for $B \ga 10^{14}$~G. \citet{Dall'Osso2015} interpret the system as having $B \sim 10^{13}$~G, only mild beaming consistent with the sinusoidal profile, and note that the low luminosity state may be due to the propeller effect. \citet{Tsygankov2016} found a bimodal luminosity distribution for M82 X-2 that they ascribe to the propeller effect and use to infer $B \sim 10^{14}$~G. The other two ULX pulsars also show strong variability that may be due to the propeller effect.  For the ULX pulsar in NGC 5907, the magnetic field inferred from the apparent luminosity is so high that object should be continually in the propeller regime.  To resolve this conundrum, \citet{Israel2016_NGC5907} suggest that the neutron star has a strong, multipolar surface field, while only the weaker dipole component is important at the magnetospheric radius.

Models have also been suggested for lower magnetic field neutron stars. \citet{Kluzniak2015} suggest that the accretion disk extends to the surface of the neutron star because the ratio of the pulse frequency derivative to the luminosity, assuming isotropic emission, implies that the accretion torque is applied at a radius comparable to the neutron star radius. In this case, the magnetic field would be weak, $B \la 10^{9}$~G, and a large fraction of the luminosity would be liberated in the disk, hence the considerations discussed below for super-Eddington black hole accretors would apply. 


Additional development of the models and their application to observational properties other than the apparent luminosity will be necessary to distinguish amongst them.  For example, the bimodal luminosity distribution of M82 X-2 appears to favor interpretation in terms of the propeller effect and a strong magnetic field. Such an effect would not be expected for a low magnetic field star in which the disk extends to the stellar surface.  The observed sinusoidal pulse shape and the energy dependence of the pulsed fraction could be compared with sufficiently developed models.  Also, future observatories with X-ray polarimetric detectors may be able to distinguish between these different proposed magnetic fields.

Given that neutron star binaries are estimated to be 10--50 times more numerous than black hole binaries \citep{Belczynski2009}, it is reasonable to ask if a large fraction of ULXs host neutron stars. In some models, beaming is predicted to be stronger by factors $\sim 10$ in neutron stars compared to black holes, which could compensate for the lower Eddington limits of neutron stars \citep{King2008,King2009}. Since the magnetic field should lose strength as it is buried by accretion, a large proportion of the ULX population may be neutron stars that do not exhibit pulsations \citep{King2016_458}. Conversely, highly super-Eddington emission from neutron stars may require such high magnetic fields that they are quite rare \citep{Tsygankov2016}.

It may be possible to make progress on this question observationally. The X-ray luminosity function of star-forming galaxies shows no break around the Eddington luminosity for a neutron star, consistent with a constant neutron star fraction even at the highest luminosities. The spectrum of NGC 7793 P13 certainly resembles other ULX spectra with its two-component form, albeit it is relatively hard \citep{Motch2014}. NGC 5907 ULX appears disk-like, similar to some other ULXs at their highest luminosities \citep{Fuerst2016}.  M82 X-2, on the other hand, may not show a turnover in its spectrum below 10 keV, unlike most ULXs \citep{Brightman2016}.  

The variability seen from the ULX pulsars is of significantly larger amplitude than seen from most ULXs and may suggest that pulsars represent a relatively small fraction of the ULX population.  Also, the one unique optical counterpart of a ULX pulsar appears different from other ULXs. The optical light, in some accretion states, is dominated by the companion star rather than the accretion disc and its wind as is typical for ULXs (Section~\ref{sec:companion}).  Hence, there are tantalizing suggestions that observational indicators other than pulsations might help us distinguish this new class.

\subsection{Black hole accretion}
\label{sec:bhsupereddmodels}

One simple way of exceeding the Eddington limit is anisotropic emission, in which case the accretion rate need not even exceed the Eddington limit for an object to appear super-Eddington if viewed along the correct sight line \citep{King2001}. Such {\it beaming\/} can take two forms: relativistic and geometric. In the former case, we view a standard X-ray binary directly along the beam of its jet (i.e. as a `microblazar'), which could boost its apparent luminosity by a factor 77 for a Lorentz factor $\gamma_{\rm j} = 5$, although this would only be seen for the 2\% of binaries that we view at the correct angle \citep{Koerding2002}. However, as we have seen, ULX spectra do not appear similar to the power-law spectra we would expect from such beamed jets. Geometric (or mild) beaming could be the result of structure in the accretion disk limiting the escape of photons such that they preferentially emerge in the directions of lowest scattering optical depth, i.e.\ along the rotational poles \citep{King2001}. The high disk scale heights necessary for this effect do not naturally occur in sub-Eddington disks; they may instead be a natural consequence of disks undergoing super-Eddington accretion.

The Eddington limit refers specifically to globally isotropic accretion; this is of course not the geometry in which disk accretion occurs.  In fact, for disk accretion the infall direction of the material (radially through the disk, towards the black hole) and the preferential direction for radiative energy release (upwards from the disk surface) are to first order perpendicular, and so the radiation should not necessarily be expected to act to limit the accretion rate on a global scale.  That disks can exceed the Eddington limit has been hypothesized since disk accretion theory was formulated \citep{Shakura1973}; this has led to a variety of models for how disks work in this supercritical accretion regime.  One early model was the so-called `Polish Doughnut' model, named for the tight, torus-like structure it predicted for the supercritical disk \citep[see][and references therein]{Abramowicz2005}.  The central regions of these axisymmetric, rotating accretion flows are optically and geometrically thick, with their height above the disk mid-plane $H(R)$, at a radius $R$, supported by radiation pressure.  \citet{Lasota2016} showed that

\begin{equation}
{{H}\over{R}} > {{3}\over{4}} {{1}\over{\eta}} {{R_{\rm S}}\over{R}} {{\dot{m}}\over{\dot{m}_{\rm Edd}}} f
\end{equation}

\noindent where $R_{\rm S} = 2GM/c^2$ is the Schwarzschild radius and $f = 1 - l_{\rm in}/l$ where $l$ and $l_{\rm in}$ are the specific angular momentum and its value at the inner boundary of the disk, respectively.  Hence, for accretion rates that exceed Eddington by factors 15 or more, the disk height $H/R \gg 5$ in its innermost regions. This can create narrow funnels of optically thick material around the rotation axis that could readily beam the X-ray emission from nearest the black hole towards an observer.

However, limitations of this model include that it only applies if the disk is radiatively efficient, and it ignores physical processes in the disk interior.  In fact, in geometrically thick super-Eddington disks, we should expect that the inflow time for material in the disk is shorter than the diffusion time for photons to reach the last scattering surface of the disk and radiate away; photons are therefore trapped in the disk, and advected into the black hole. Models that take this effect into account are generally termed `slim disk' models \citep{Abramowicz1988,Sadowski2009}, that are characterized by low radiative efficiencies and, as Equation (3) is modified in proportion to the radiative efficiency, hence $H/R \la 1$.  Although much work has focussed on slim disks occurring at around the Eddington rate \citep{Abramowicz2005}, recent studies have shown that advection acts to limit the disk height and, as a result, degree of beaming at all super-Eddington rates, such that we should always expect the disks to be slim rather than very geometrically thick \citep{Lasota2016,Wielgus2016}. Indeed, a further set of models predict that super-Eddington accretion and, in particular, radiation release is possible from thin disks if extreme accretion rates exacerbate their internal density inhomogeneities and lead to the non-linear development of photon bubbles. This results in the disk becoming somewhat porous to radiation from the interior of the disk, permitting it to escape, hence the epithet `leaky disks' \citep{Begelman2002}.

\begin{figure}[tb]
\includegraphics[width=4in]{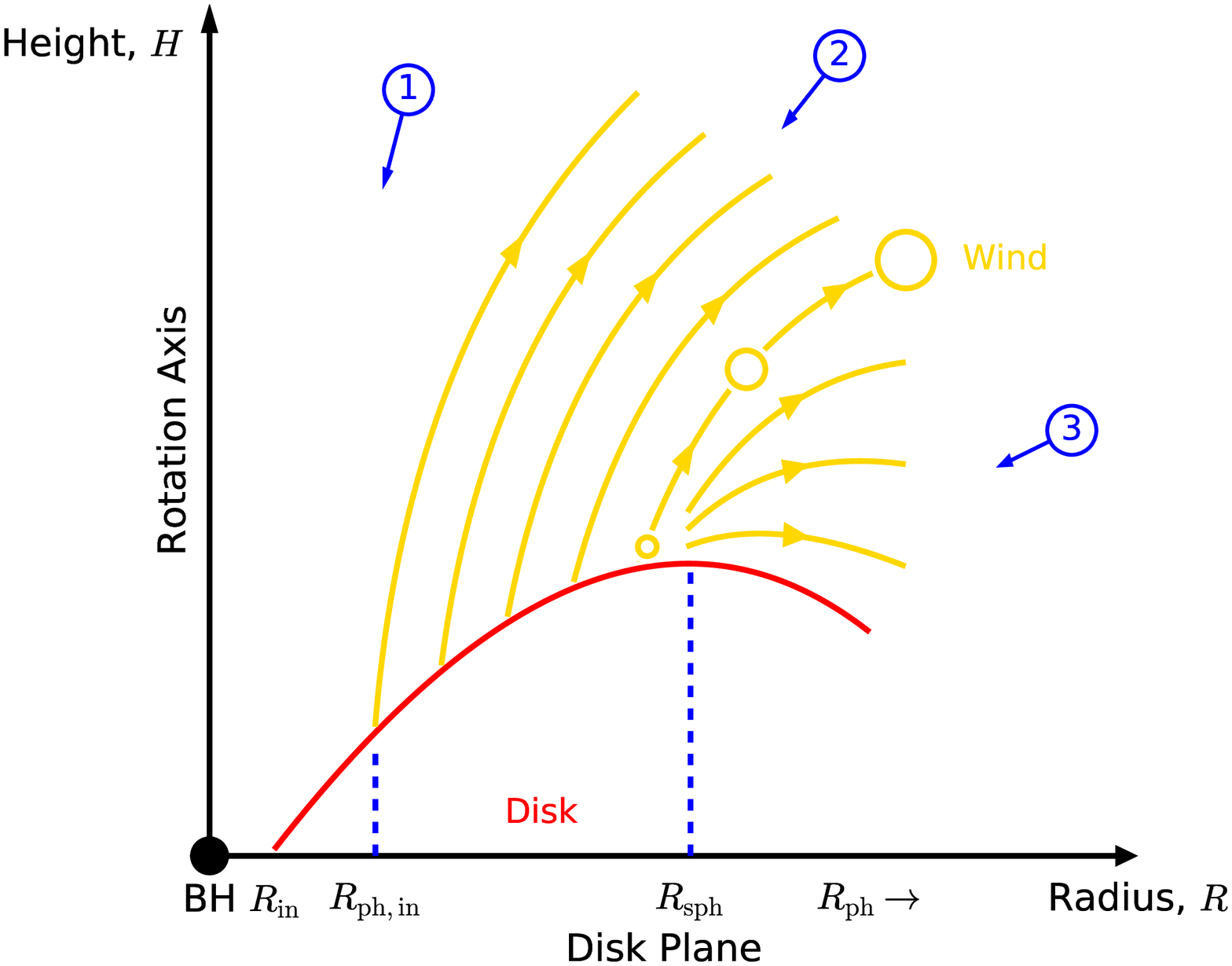}
\caption{Diagram of a super-critical accretion flow.  The accretion disc (in red) becomes geometrically thick at $R_{\rm sph}$, and within this radius a massive, optically-thick outflow is blown from the top of the disk by radiation pressure (the wind, in yellow), which persists down to $R_{\rm ph,in}$.  The wind is clumpy in nature, but as it moves away from the disk the wind material diffuses perpendicular to its path of motion, illustrated by the expanding blob blown from $R_{\rm sph}$.  However, the wind remains optically thick well beyond the region shown, with the outer photosphere (region to $R_{\rm ph}$) extending well over the disk. This geometry results in harder spectra for lines of sight close to the rotation axis of the black hole (1), softer spectra as the line of sight crosses the wind (2), and ultrasoft or UV-dominated spectra for lines of sight that view only the outer photosphere (3).}
\label{fig:supEddmodel}
\end{figure}

A further characteristic of super-Eddington accretion is that the Eddington limit always applies locally; thus excess material will be driven from the disk, in the form of a massive wind. This becomes important when the mass transfer rate exceeds $(9/4) \dot{m}_{\rm Edd}$ \citep{Shakura1973}.  In this scenario the outer parts of the accretion disk remain geometrically thin down to a certain radius $R_{\rm sph}$, the spherization radius, where $R_{\rm sph} = (\dot{m}/\dot{m}_{\rm Edd}) R_{\rm in}$ in units of the inner disk radius $R_{\rm in}$. Within this radius the disk becomes geometrically thick as radiation pressure inflates it, and any excess material is removed from the disk's upper layers in the form of a radiation-pressure driven wind. This wind will be very massive and will remain optically thick as it moves away from the disk, thereby providing collimation of the X-ray emission from the central regions. Sufficient material is lost in the wind that (in the absence of advection) the accretion rate at the inner edge of the disk should be equal to $\dot{m}_{\rm Edd}$. In such a model, we can express the apparent luminosity of a ULX as

\begin{equation}
L \simeq {{L_{\rm Edd}}\over{b}} \left[1+\ln \left({{\dot{m}}\over{\dot{m}_{\rm Edd}}}\right)\right]
\label{eqn:beaming_logmdot}
\end{equation}

\noindent where $b$ is the beaming factor \citep{King2008}, and this varies with the accretion rate (in Eddington units) as $b \propto (\dot{m}/\dot{m}_{\rm Edd})^{-2}$ \citep{King2009}.  This model of a geometrically thick disk and a radiation-driven wind derives directly from the supercritical disk model first described by \citet[][section IV]{Shakura1973}, and more recently expanded on by \citet{Poutanen2007} and \citet{Middleton2015_447}.  Here we briefly describe its structure, based on these works, and present a schematic in Figure~\ref{fig:supEddmodel}.

The structure of a supercritical flow can be divided into 3 main zones, demarcated by three different radii: $R_{\rm ph,in}$, the inner edge of the photosphere at which the optical depth of the photosphere drops to unity; $R_{\rm sph}$; and $R_{\rm ph}$, the outer edge of the photosphere at which the optical depth of the photosphere drops to unity.  These zones are:
\begin{itemize}
\item
{\bf Zone A: $R < R_{\rm ph,in}$.}  At the smallest disk radii the wind is optically thin to electron scattering in the vertical direction, and the X-ray emission resembles a distorted, hot accretion disk.
\item
{\bf Zone B: $R_{\rm ph,in} < R < R_{\rm sph}$.}  The wind is opaque in this region, becoming effectively a vertical extension of the inflow.  Advection is important in this material, and so it has a $R^{-1/2}$ temperature profile, with the radiation escaping at a radius roughly double that of where it is produced as it is advected by the wind.
\item
{\bf Zone C $R_{\rm sph} < R < R_{\rm ph}$.}  The optical depth of the wind blown into this region falls as $1/r$, and so the radiation emerges at approximately the radius it is produced.  The temperature profile appears as the classic disk $R^{-3/4}$ profile, with roughly the Eddington luminosity generated in this region.
\end{itemize}

Crucially, the geometry of the supercritical accretion scenario (illustrated in Fig.~\ref{fig:supEddmodel}) means that different zones will be visible at different viewing angles for the system, hence its predicted appearance varies as a function of inclination to the line of sight. For example, the line of sight to the central regions of the accretion flow (1 in Fig.~\ref{fig:supEddmodel}) will be dominated by the inner disk, with this emission enhanced by beaming up the evacuated funnel.  At intermediate angles (2), the view will pass through the wind, and so be dominated by the soft, thermal component with a temperature similar to that at the spherization radius.  At the largest inclination angles (3) the view will be of the outer photosphere, that will be much cooler to the point that its emission will be predominantly in the UV regime. Changes in the accretion rate affect the mass flow entering the wind and therefore the wind scale height, thus changing the appearance at fixed inclination. 

The validity of the main features of this model have been confirmed via numerical simulations. These require the use of advanced radiation-magnetohydrodynamics (RMHD), or general relativistic RMHD codes, to fully simulate the conditions in accretion flows \citep[e.g.][]{Ohsuga2011,Jiang2014,Sadowski2015}. These models confirm the presence of a large scale height inflow, a powerful outflow, and a central evacuated funnel. However, they show some differences to the analytical predictions. \citet{Takeuchi2013} find a clumpy wind that they suggest is due to Rayleigh-Taylor instabilities. In addition, \citet{Jiang2014} and \citet{Sadowski2016} find that vertical advection of radiation caused by magnetic buoyancy increases radiation escape from the disk producing an efficiency not too dissimilar to standard disks, $\eta \sim 0.03 - 0.08$, although the location of the bulk of the radiation release differs strongly between these models.  Interestingly, these models do not display any hint of photon bubbles \citep{Begelman2002}, likely due to the turbulent nature of the disk interiors.

\noindent\begin{textbox}[tb]\section{Super-Eddington Spectral Regimes}
{\bf Ultraluminous (UL) --} The spectra are distinctly two-component. The soft thermal component is thought to arise from the inner photosphere within the outflow and can be modeled with a simple blackbody or a multicolor black body with a temperature of 0.1--0.3~keV. The hard component is thought to arise from the hot, inner regions of the accretion disk and can be modeled as a DBB or optically thick corona with temperatures of $1.5-3$~keV. Hard UL spectra would correspond to low inclination sight lines (near \#1 in Fig~\ref{fig:supEddmodel}) where the inner disk is prominent. Soft UL spectra would arise from moderation inclination (near \#2 in the figure) where the view of the inner disk is reduced and that of the photosphere is increased.
\newline
{\bf Supersoft ultraluminous (SSUL) --} The spectra may correspond to large inclinations (near \#3 in the figure). The inner disk would then be obscured and the spectrum dominated by emission from the outer and cooler photosphere. SSUL spectra may also arise when the accretion rate increases to the extent that the outflow covers the disk at most inclinations.
\newline
{\bf Broadened disk (BD) --} The spectra are thought to be dominated by the accretion-disk, but with the disk structure modified from the standard thin disk due to the high accretion rate. For the bulk of the BD population near $10^{39} \rm ~erg~s^{-1}$ this could correspond to near Eddington accretion rates, but the much more luminous examples may need a different explanation, for example they may be UL objects viewed down the funnel where the beamed inner disk emission dominates the X-ray spectrum.
\newline
\end{textbox}



X-ray spectra are indicative of the processes within the inner accretion flow, and so can tell us much about whether the flows are behaving as we would expect for a super-Eddington flow. Here, we attempt to relate models of super-Eddington accretion to the observational results on ULXs presented above. The sidebar ``Super-Eddington Spectral Regimes'' describes the phenomenological classification of the ultraluminous spectral regimes presented in Section~\ref{sec:xrayspectra} and graphically in Fig.~\ref{fig:ULXspecs} in terms of a super-critical accretion flow as illustrated in Fig.~\ref{fig:supEddmodel}.  We note that the similarity of the ULX pulsar spectra to the spectra discussed here raises a serious caveat in the interpretation of other ULX spectra in terms of black hole accretion.

Some ULX spectra can be interpreted in terms of the emission expected from `slim disk' models, where advection within the disk becomes a dominant process.  This should change the relationship between luminosity and temperature, flattening it such that $n < 4$ in the classic $L \propto T^n$ relationship for disk luminosity $L$ and temperature $T$; this is seen in ULXs \citep{Watarai2001}. Similarly, the radial emission profile of the disk temperature is $T \propto R^{-p}$ at a radius $R$, where $p = 0.75$ is expected for a standard thin accretion disk, and $p = 0.5$ for a slim disk.  Model fits with $p$ as a free parameter show that ULXs tend to have $p \approx 0.6$; however this does not constitute a good fit to all high quality ULX spectra, with slim disks tending to fit better to BD spectra (Section~\ref{sec:contspec}). In addition, a study of the highest quality data from a BD ULX, M33 X-8, demonstrates that it does not behave as per the expectations of a simple slim disk \citep{Middleton2011_417}, and a study of a transient ULX in M31 shows that advection does not alter its spectrum substantially from that predicted in a standard disk \citep{Straub2013}.  Nevertheless, this remains a possible physical solution for many ULXs with BD spectra, for example a variable ULX in M83 \citep{Soria2015}.


The two-component spectra of brighter ULXs have been physically interpreted via supercritical accretion models, with the soft component being emission from the inner photosphere within the outflowing wind \citep{King2003,Poutanen2007} and the hard component being the hot, inner disk emission \citep{Middleton2011_411,Kajava2012}. This differs from the original interpretation of the two-component UL spectra discussed above. The luminosity of the soft/wind component is then predicted to decrease with increasing temperature \citep{Poutanen2007}, in strong contrast with the expected behavior of a disk. As discussed in Section~\ref{sec:spec_var}, the soft component of ULX spectra fitted with two-component models generally show decreasing luminosity, both total and in the soft component, with increasing temperature \citep{Feng2007}, with a relationship $L \propto T^{-3.5}$ \citep{Kajava2009}. 

In the supercritical model, the observational characteristics of ULXs should be a function of the viewing angle, with harder spectra for lines of sight close to the rotation axis of the black hole (low inclination, \#1 in Fig.~\ref{fig:supEddmodel}), softer spectra as the line of sight crosses the wind (moderate inclination, \#2 in Fig.~\ref{fig:supEddmodel}), and ultrasoft or UV-dominated spectra for lines of sight that view only the outer photosphere (high inclination, \#3 in Fig.~\ref{fig:supEddmodel}). For the latter, the observed X-ray luminosity should be low as the hot disk is hidden. These three lines of sight match the hard UL regime, the soft UL regime, and the SSUL regime, respectively. Furthermore, the wind height is predicted to increase with the accretion rate increasing the beaming of emission in the central funnel. Hence, the observer's view of the disk versus wind emission will change with accretion rate. At low inclinations where the ULX is viewed down the funnel, the beaming of the hard component increases as $\dot{m}^2$ while the soft component luminosity increases as $\dot{m}$, so the spectrum gets harder with increasing $\dot{m}$ and becomes dominated by the hard component alone \citep{Middleton2015_447}. At moderate inclination angles, one would see the soft (wind) emission increase relative to the hard (disk) emission as the luminosity increases. \citet{Sutton2013_435} see two ULXs that clearly demonstrate this latter behavior.

Accretion disks are known to precess, which could change our observed inclination angle; this would manifest as spectral degeneracy with luminosity. Spectral degeneracy has been seen in ULXs (Section~\ref{sec:spec_var}). \citet{Luangtip2016} show that the spectral variability and degeneracy of Ho IX X-1 fit into the patterns expected for variation of both accretion rate (soft component $\propto \dot{m}$ and hard component $\propto \dot{m}^2$ for an object viewed into its funnel) and inclination angle.

The patterns of flux variability with spectral shape in ULXs also support a super-Eddington interpretation. In particular, \citet{Sutton2013_435} show that high levels of rms variability are not seen in hard UL or (most) BD observations; yet they are common for soft UL spectra. This variability is much stronger above 1 keV than below and the spectrum of the variable component matches that of the hard component (Section~\ref{sec:lowfreqvar}). This is best interpreted as extrinsic variability, imprinted on the hard component in the ULX spectra, by its passage through a clumpy wind \citep{Middleton2011_411,Takeuchi2013}; the softness of the spectra support this by implying that our line-of-sight must at least graze the wind. However, \citet{Feng2016} note that the timing properties of the soft component in SSULXs are different from those seen in soft UL sources and suggest that the accretion rate may play a more important role than the viewing angle. SSUL spectra may also arise when the accretion rate increases to the extent that the outflow covers the disk at most inclinations.

However, there are some observations that remain unexplained in this scenario.  Most notably the mHz QPOs seen in some ULXs  (Section~\ref{sec:lowfreqvar}) and the soft lags recently reported in two objects \citep{Hernandez-Garcia2015} are unexplained. Also, the detection of an additional, harder component of the X-rays spectrum above 10~keV in some {\it NuSTAR\/} observations (Section~\ref{sec:contspec}) is currently unexplained within the context of supercritical accretion.

\begin{figure}[tb]
\includegraphics[width=4in]{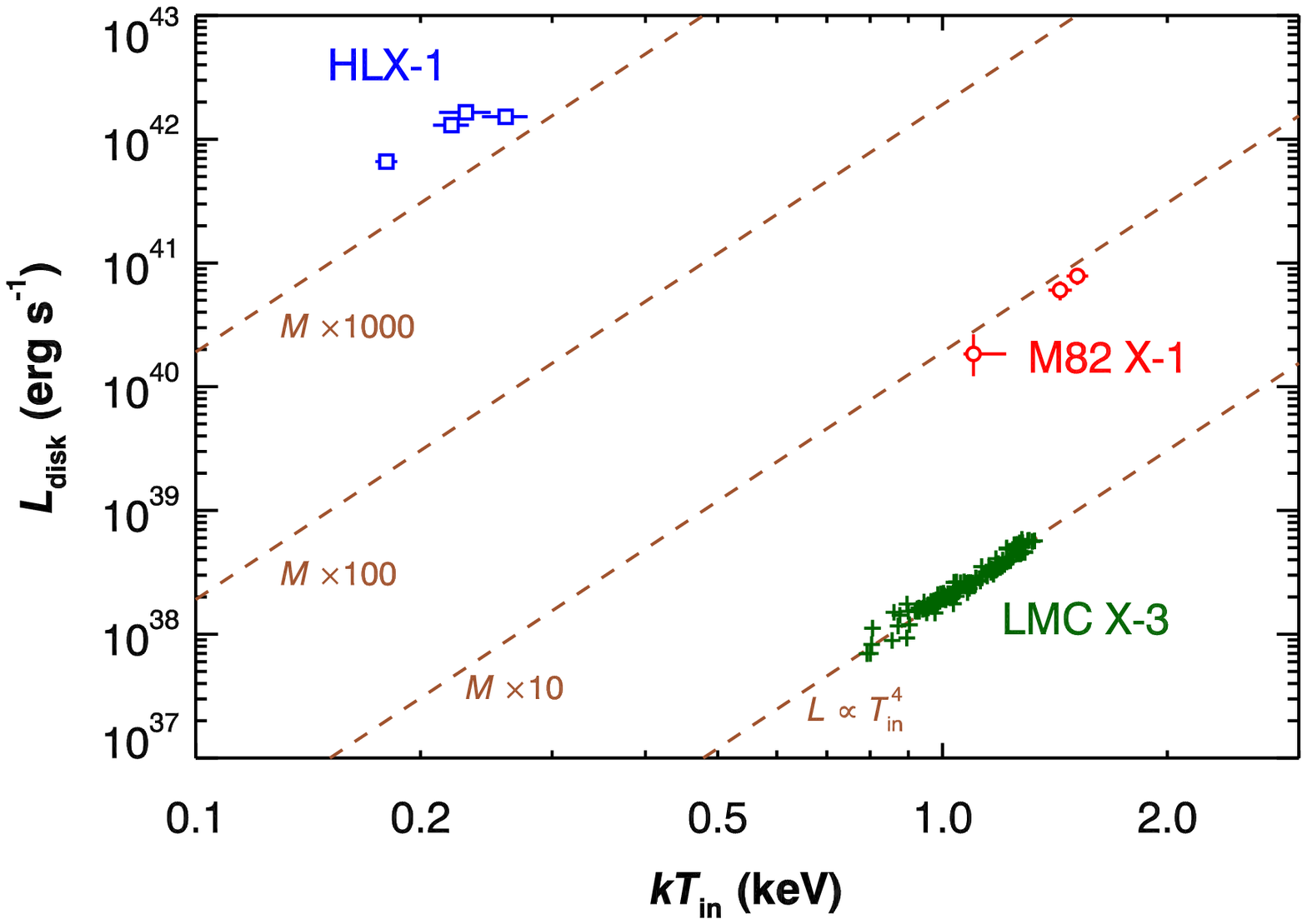}
\caption{Bolometric disk luminosity versus the disk inner temperature for the BHB LMC X-3 \citep[data from][]{Gierlinski2004}, and the two IMBH candidates, M82 X-1 \citep[data from][]{Feng2009} and ESO 243-49 HLX-1 \citep[data from][]{Servillat2011}. The lowest dashed line represents a fit of the LMC X-3 data to a $L_{\rm disk} \propto T_{\rm in}^4$ relation. Each higher dashed line represents a factor of 100 increase in luminosity corresponding to a mass increase of a factor of 10 at fixed Eddington ratio, black hole spin, and disk inclination.}
\label{fig:lt}
\end{figure}

\section{INTERMEDIATE-MASS BLACK HOLE CANDIDATES}
\label{sec:imbh}

In the mass spectrum of black holes, two populations have been firmly identified, stellar remnant black holes with masses $\sim 10 M_\odot$ that are the remnants of individual stars and supermassive black holes (SMBHs) in the nuclei of galaxies with masses of $10^6 - 10^9 M_{\odot}$. Stellar black holes have been observed with masses up to 20--30~$M_{\odot}$ \citep{Prestwich2007} This result, for IC 10 X-1, has been disputed by \citet{Laycock2015}; however indisputable evidence for black holes of this mass came from the first detection of a gravitational wave source (Section~\ref{sec:gravwave})., but ones as massive as $\sim 80 M_{\odot}$ might be formed from low metallicity stars \citep{Belczynski2010}. Stellar black holes may also increase their mass via accretion or mergers. The low mass end of the SMBH distribution is an area of active research and extends at least as low as $3 \times 10^5 M_{\odot}$ \citep{Peterson2005}. The masses of SMBHs are correlated with the properties of the host galaxy bulge, suggesting linked evolution, but the formation mechanism is unknown.

The existence of black holes with masses in between ($10^2 - 10^5 M_\odot$), the so-called intermediate mass black holes (IMBHs), is an open question. The high end of the IMBH range is sometimes taken to overlap with the low end of the SMBH range, but here we keep the two classes distinct and consider only objects thought to be accreting from a companion star rather than the interstellar medium. The discovery of an IMBH that is not a low-mass SMBH while well above the maximum possible stellar black hole mass would require a novel mechanism for its formation. Such objects may have been important in the formation of SMBHs \citep[\textit{cf}.][see Section~\ref{sec:supereddagn}]{Volonteri2010}.

There is a break in the XLF of star-forming galaxies at (1--2)$\times 10^{40} \rm \, erg \, s^{-1}$ (Section~\ref{sec:xlf}). Objects at higher luminosities, HLXs, appear to be a separate class \citep{Swartz2011}. HLXs are candidate IMBHs because their extremely high luminosity is difficult to explain by super-Eddington accretion onto stellar mass black holes. We note that HLXs are rare in the local universe, see searches based on {\it Chandra} \citep{Gao2003,Gong2016} and {\it XMM-Newton} \citep{Sutton2012,Zolotukhin2016} data. Contamination by foreground stars and background AGN is severe for HLX catalogs, reaching $\sim$70\% \citep{Zolotukhin2016}; optical observations are needed to remove interlopers \citep[{\it e.g.},][]{Sutton2015}.

\subsection{ESO 243-49 HLX-1}
\label{sec:iso243-49}

ESO 243-49 HLX-1 has an extremely high luminosity, peaked above $10^{42} \rm \, erg \, s^{-1}$ and is well removed from the nucleus of the host galaxy ESO 243-49 \citep{Farrell2009}. The source makes transitions from the hard state to the thermal state, as seen in GBHBs. \citet{Servillat2011} found that the disk emission follows the $L_{\rm disk} \propto T_{\rm in}^4$ relation, indicative of a standard accretion disk (Section~\ref{sec:gbhb}), see Fig.~\ref{fig:lt}. The lowest dashed line in the figure is a fit of data for the BHB LMC X-3 to a $L_{\rm disk} \propto T_{\rm in}^4$ relation \citep{Gierlinski2004}. The black hole in LMC X-3 has a mass of $7.0 \pm 0.6 M_\odot$ \citep{Orosz2014}, a low spin consistent with zero \citep{Steiner2014}, and the luminosities in the plot range from 0.07 to $0.6 \, L_{\rm Edd}$. The disk luminosity at fixed Eddington ratio scales as $L_{\rm disk} \propto \alpha^2 M^2 T_{\rm in}^4$ where $\alpha =1$ for a non-rotating black hole and $\alpha = 1/6$ for a maximally spinning one and $M$ is the black hole mass \citep{Makishima2000}. If ESO 243-49 HLX-1 was in the thermal state during the observations, then its mass is roughly $2000 \times$ that of LMC X-3. However, the disk luminosity can vary by a factor of 36 between a non-spinning and a maximally spinning black hole given the same mass and disk inner temperature, hence larger masses are allowed. The unknown inclination angle ($i$) contributes a factor of $1/\cos i$ to the luminosity.

Improved accretion disk models give similar results. \citet{Davis2011} used a fully relativistic thin disk model and obtained a lower bound of the black hole mass as 3,000~$M_\odot$. \citet{Godet2012} implemented a simplified slim disk model that took into account radial advection at high accretion rates and derived a black hole mass of $\sim 2 \times 10^4 M_\odot$ assuming a non-spinning black hole with a face-on disk. They also confirmed the $4^{\rm th}$ power relation between disk luminosity and temperature. \citet{Straub2014} applied a fully relativistic slim disk model and constrained the mass to be between 6,000 and 200,000~$M_{\odot}$; the wide range is due to the unknown spin. Radio emission from HLX-1 has been detected \citep{Webb2012,Cseh2015}, but the mass limits are not constraining  (Section~\ref{sec:compactradio}).

The optical counterpart ESO 243-49 HLX-1 emits an H$\alpha$ emission line establishing the object as physically associated with ESO 243-49 \citep{Wiersema2010} and has a magnitude consistent with a massive star cluster \citep{Soria2010}. Variability in the optical flux suggests that part arises from the accretion flow \citep{Soria2013,Webb2014}. The high velocity offset of the H$\alpha$ line, $420 \rm \, km \, s^{-1}$ relative to the galactic nucleus and $270 \rm \, km \, s^{-1}$ relative to the stellar rotational velocity near its location, suggests that the surrounding star cluster is the remnant of a tidally stripped dwarf satellite galaxy \citep{Soria2013}. In this case, ESO 243-49 HLX-1 may represent an object at the low end of the SMBH distribution.


\subsection{M82 X-1}
\label{sec:m82x1}

M82 X-1 exhibits a peak luminosity close to $10^{41}$~erg~s$^{-1}$ and is another promising IMBH candidate on the basis of two pieces of evidence. One is the detection of a pair of HF-QPOs as discussed in Section~\ref{sec:timing}. The high coherence and 3:2 frequency ratio measured with 2\% accuracy of these QPOs is strong evidence in favor of interpreting them as HF-QPOs from a black hole with a mass of $428 \pm 105 M_\odot$ \citep{Pasham2014}. 

The other evidence comes from correlated spectral and low frequency timing variability. In 2008/2009, three joint {\it Chandra} and {\it XMM-Newton} observations failed to detect low frequency timing noise; the {\it XMM-Newton} data placed upper limits on the fractional rms (6\%-7\% at 3$\sigma$ confidence) well below previous detections \citep{Feng2010}. The energy spectrum also changed shape from a straight power-law to having significant curvature\footnote{A reliable X-ray spectrum for M82 X-1 can be obtained only with moderately offset {\it Chandra} ACIS observations plus sub-array readout, such that source confusion can be avoided and the CCD pile-up effect can be reduced to an acceptable level. Spectra for M82 X-1 measured with {\it Chandra} in a different configuration or with other telescopes should be treated with caution.}. The spectra are best fitted by a DBB model with the inner disk temperature varying from 1.1 to 1.5 keV and following $L_{\rm disk} \propto T_{\rm in}^4$ relation, see Fig.~\ref{fig:lt}, consistent with accretion disk emission. Disk temperature scales as $T_{\rm in} \propto \alpha^{-1/2} \eta^{1/4} M^{-1/4}$ where $\eta = L_{\rm disk}/L_{\rm Edd}$. Thus, the higher disk temperature at higher luminosity, compared with LMC X-3, suggests a rotating black hole with $\alpha < 1$. Assuming high spin ($a^* > 0.93$), \citet{Feng2010} derived a black hole mass of 200-800 $M_\odot$. The {\it Chandra} spectra have a limited bandpass due to high absorption and {\it Chandra}'s poor high energy response. \citet{Brightman2016} analyzed simultaneous {\it NuSTAR} and {\it Swift} spectra of M82 that have good high energy coverage, but potential issues with source confusion, with a slim disk model. They find a mass of 20--118~$M_{\odot}$ for a non-rotating black hole and 105--573~$M_{\odot}$ for a maximally-rotating black hole. The latter range is shifted lower than, but overlaps with, that from \citet{Feng2010} and that derived from the HF-QPOs. 

M82 X-1 lies near the super star cluster MGG 11. A possible mechanism for the formation of IMBHs is stellar collisions in the cores of dense stellar clusters. \citet{Portegies-Zwart2004} found that the extremely compact size of MGG 11, its half light radius is only 1.2~pc, causes massive stars to rapidly sink to the cluster center via dynamical friction and makes the cluster a good candidate for the production of an IMBH.

\subsection{Other candidates}

\citet{Sutton2012} examined 8 HLXs in {\it XMM-Newton} data and found that their spectra are well modeled by an absorbed power-law with $\Gamma \sim 1.7$, which is harder than typical ULXs. Some show evidence for strong variability, with rms of 10--20\% on timescales of 0.2--2 ks. These properties are consistent with the hard state, suggesting that the sources are sub-Eddington and that their high luminosities indicate high black hole masses. One notable object in the sample is NGC 2276-3c from which \citet{Mezcua2015} reported a compact jet, described in Section~\ref{sec:compactradio}. If confirmed, the radio jet would verify that the HLX is in the hard state and establish it as an IMBH.


\section{IMPLICATIONS}

\subsection{Super-Eddington Accretion in Active Galactic Nuclei}
\label{sec:supereddagn}

It is now thought that the supermassive black holes we see in most galactic nuclei have grown over time from seed black holes, created when the Universe was only a small fraction of its current age, via accretion and mergers.  The mass and creation mechanisms for these seeds are the subject of much debate, with possible seed masses ranging from large stellar mass black holes, $\sim 40 M_{\odot}$, to the top end of the IMBH regime, $\sim 10^5 M_{\odot}$ \citep{Volonteri2010}.  However, the presence of QSOs containing very massive black holes at high redshifts \citep[e.g. $2 \times 10^9 M_{\odot}$ at $z \approx 7$, a mere 0.78 Gyr after the big bang,][]{Mortlock2011} means that the growth of the seeds cannot be Eddington-limited; the only way to get such massive objects that early in cosmic history is through super-Eddington accretion.  It is also becoming apparent that there are some super-Eddington AGN in the local Universe, commonly found amongst those AGN classified as narrow-line Seyfert 1 galaxies \citep[e.g. RX J1140.1+307,][]{Jin2016}.  Super-Eddington accretion is therefore not solely the domain of stellar-mass objects.

The physics of super-Eddington accretion onto supermassive black holes is thought to broadly resemble that seen in ULXs, with one important caveat: the accretion is generally from an ISM, rather than a dense gravitationally-bound object such as a star.  This means that the accretion flow is more susceptible to being impeded by the radiation and/or mechanical pressure released in accretion.  For example, \citet{Sakurai2016} show that the large-scale inflow towards a super-Eddington AGN is only stable for luminosities below a certain threshold; above this, the inflow becomes strongly episodic.  The physics of the accretion disk itself is very similar to ULXs, with recent modeling showing that slim disk solutions permit the growth rates to increase significantly above that expected from Eddington-limited accretion during short (10 kyr - 10 Myr), heavily-obscured accretion episodes \citep{Volonteri2015}.  Such episodes should also drive strong outflows, as in ULXs, that should provide mechanical feedback that contributes to limiting the accretion and so affects the black hole/galaxy spheroid mass relations we see at the current epoch.  \citet{King2016_455} calculate the strength of this wind for hyper-Eddington accretion (accretion rates $> 10^3$ times Eddington), and note that this would lead to significantly smaller black holes than are seen in bulges today as material is swept away and not accreted; hyper-Eddington accretion must therefore be very rare in AGN, with their super-Eddington epochs limited to luminosities close to Eddington.

\subsection{X-Ray Binaries in Early Galaxies}
\label{sec:earlygalaxies}

X-ray binaries likely dominated the X-ray emission in the early universe and contributed to the heating of the intergalactic medium (IGM) during the epoch of reionization when the IGM changed from being cold and neutral to being warm and ionized \citep{Jeon2014}. Detection of X-ray binaries in the early universe is not currently possible. One observational avenue to derive the properties of early X-ray binaries is the study of local analogs to high redshift galaxies. The early universe was highly metal deficient, even at $z = 6$ the average metallicity is $\sim 0.1 \, Z/Z_{\odot}$ \citep{Savaglio2006}. Metallicity appears to have a strong impact on the formation and properties of X-ray binaries, particularly in the high luminosity range, hence study of ULXs is important to inform our understanding of the early universe.

Studies of X-ray binaries and ULXs in star-forming galaxies show that their production, relative to the star formation rate (SFR), is enhanced at sub-solar metallicities \citep{Mapelli2010,Basu-Zych2013}. These trends are enhanced at very low metalliticies, $Z/Z_{\odot} < 0.1$, with an increase, relative to near-solar-metallicity galaxies, of a factor of $7 \pm 3$ in the number of ULXs \citep{Prestwich2013} and $11.5 \pm 2.7$ in the total galactic X-ray luminosity \citep{Brorby2014}. Hence X-ray binary production was likely enhanced in the early universe. \citet{Brorby2016} suggested a `fundamental plane' relation linking the total X-ray luminosity of a galaxy to its star formation rate and metallicity with a scatter of 0.25~dex that has been used in simulations of heating by X-ray binaries in the early universe \citep{Mirocha2016}. Because the total luminosity is dominated by ULXs, the ULX spectral shape determines the penetration of the X-rays into the IGM and, thus, the morphology of X-ray heating having implications for the detection of 21-cm hydrogen hyperfine radiation from the epoch of reionization 
\citep{Kaaret2014}.

Population synthesis calculations of binary formation and evolution also show enhanced X-ray emission and suggest physical causes.  The decreased mass transfer due to radiative-driven winds from the companion stars increases the number of binary systems evolving into Roche-lobe overflow systems \citep{Linden2010}. The decreased mass loss increases the maximum mass of black holes formed at the end of stellar evolution, reaching as high as $80 M_{\odot}$ for $Z/Z_{\odot} = 0.01$ versus $\sim 20 M_{\odot}$ for $Z/Z_{\odot} \sim 1$ \citep{Belczynski2010}.

Lyman $\alpha$ and Lyman continuum radiation from massive stars, that are copiously produced in starburst regions in early galaxies, is thought to have powered reionization of the IGM. However, the fraction of the radiation that escapes from galaxies is low because Lyman continuum and line emission is absorbed by dust and Ly$\alpha$ is resonantly scattered by neutral hydrogen. Some source of feedback is required to blow neutral gas and dust away from the starburst to allow the Lyman emission to escape \citep[e.g.][]{Orsi2012}. As summarized in Section~\ref{sec:supereddevidence}, there is strong evidence for powerful outflows from ULXs. \citet{Prestwich2015} identified two candidate ULXs in Haro 11, one of the few confirmed Lyman continuum emitting galaxies in local universe. They suggested that outflows from the ULXs blow neutral material away from the starburst regions where the ULXs are located, allowing the Lyman radiation produced by the massive stars in the starburst to escape. If the correlation between ULXs and Lyman continuum emitting galaxies is confirmed with a larger sample, this may indicate that ULXs had an important role in the reionization of the IGM.

\subsection{Gravitational Wave Sources}
\label{sec:gravwave}

The first gravitational wave event (GW150914) discovered by the Laser Interferometer Gravitational-Wave Observatory (LIGO) was produced by a pair of black holes with masses of $36^{+5}_{-4} M_{\odot}$ and $29 \pm 4 M_{\odot}$ \citep{LIGO2016}. These are larger than any known stellar black hole in the Milky Way \citep{Remillard2006}, but within the range suggested for some ULXs and theoretically predicted to be produced in low metallicity environments.

\citet{Belczynski2016} performed population synthesis simulations that suggest that the binary progenitor of GW150914 was formed in an environment with a metallicity less than 10\%. A possible binary evolution leading to the observed merger begins with massive stars ($\sim 60$ and $\sim 100$ $M_{\odot}$) that evolve through a common envelope phase into an X-ray binary consisting of a He star and a $36 M_{\odot}$ black hole. ULX binaries are formed through similar channels and offer a means to test the simulations, potentially helping to determine whether binary evolution or dynamical interactions in clusters is the dominant production mode for gravitational wave event progenitors.





\section*{DISCLOSURE STATEMENT}
The authors are not aware of any affiliations, memberships, funding, or financial holdings that might be perceived as affecting the objectivity of this review. 

\section*{ACKNOWLEDGMENTS}
We thank Matthew Middleton and Matthew Brorby for useful comments and Marat Gilfanov, Lian Tao, Marek Gierli\'{n}ski, and Dong-Woo Kim for supplying data for the figures. PK acknowledges funding support from NASA. HF acknowledges funding support from the National Natural Science Foundation of China under grant No.\ 11633003, the National Program on Key Research and Development Project under grant No.\ 2016YFA040080X, and the Tsinghua University Initiative Scientific Research Program. TPR acknowledges support from STFC as part of the consolidated grant ST/L00075X/1. 

%


\end{document}